\documentclass[12pt]{iopart}
\newcommand{\vc}[1]{\mbox{\boldmath\/${#1}$}}
\usepackage{graphicx}
\usepackage{iopams}

\begin{document}

\title[High order fluid model for streamer discharges]{High order fluid model for streamer discharges: \\ I. Derivation of model and transport data}

\author{S. Dujko$^{1,2}$, A.H. Markosyan$^{1}$, R.D. White$^{3}$ and U. Ebert$^{1,4}$}
\address{$^1$ Centrum Wiskunde \& Informatica (CWI), P.O. Box 94079, 1090 GB Amsterdam, The Netherlands}
\address{$^2$ Institute of Physics, University of Belgrade, P.O. Box 68, 11080 Zemun, Belgrade, Serbia}
\address{$^3$ ARC Centre for Antimatter-Matter Studies, School of Engineering and Physical Sciences, James Cook University, Townsville 4810, Australia}
\address{$^4$ Department of Applied Physics, Eindhoven University of Technology, P.O. Box 513, 5600 MB Eindhoven, The Netherlands}

\ead{S.Dujko@cwi.nl}

\begin{abstract}
Streamer discharges pose basic problems in plasma physics, as they are very transient, far from equilibrium and have high ionization density gradients; they appear in diverse areas of science and technology. The present paper focuses on the derivation of a high order fluid model for streamers. Using momentum transfer theory, the fluid equations are obtained as velocity moments of the Boltzmann equation; they are closed in the local mean energy approximation and coupled to the Poisson equation for the space charge generated electric field. The high order tensor in the energy flux equation is approximated by the product of two lower order moments to close the system. The average collision frequencies for momentum and energy transfer in elastic and inelastic collisions for electrons in molecular nitrogen are calculated from a multi term Boltzmann equation solution. We then discuss, in particular,  (1) the correct implementation of transport data in streamer models; (2) the accuracy of the two term approximation for solving Boltzmann's equation in the context of streamer studies; and (3) the evaluation of the mean-energy-dependent collision rates for electrons required as an input in the high order fluid model. In the second paper in this sequence, we will discuss the solutions of the high order fluid model for streamers, based on model and input data derived in the present paper.
\end{abstract}

\submitto{\JPD}

\section{Introduction}
\label{sec1}

Streamers are rapidly growing filaments of weakly ionized plasma, whose dynamics are controlled by highly localized space charge regions and steep plasma density gradients. The dynamics of the streamer ionization front are governed by electron dynamics in electric fields above the break-down value; therefore the plasma is very far from equilibrium, and the neutral gas stays cold while the ionization front passes.

Streamers occur widely in pulsed discharges, both in nature and in technology. As their size scales with inverse gas density, streamers occur in the limited volumes of plasma technological devices mostly at high gas densities, while planetary atmospheres also can host huge discharges at very low gas densities; so-called sprite discharges~\cite{Pasko2006, EbertNLLBV2010, Luque2012} are streamers, they exist at altitudes of up to 90 km in our atmosphere, i.e., at pressures down to well below 10~$\mu$bar. For a cluster issue on streamers, sprites and lightning, we refer to~\cite{JPDCI2008} and the 19 original articles therein, discussing common issues of streamer dynamics in atmospheric discharges and plasma technology, including discharge evolution and subsequent chemical reactions. Streamers are a key element in gas processing in so-called corona reactors~\cite{Winands2008, Heesch2008} as well as in high voltage technology~\cite{Seeger2009}; they are used for the treatment of polluted gases~\cite{Veldhuizen2000} and water~\cite{GrabowskiVPR2006} or for plasma enhanced vapor deposition \cite{BabayanJTPSH1998}. The plasma bullets observed in plasma jets in noble gases have now been identified with streamers~\cite{JiangJC2009, Naidis2010, Naidis2011, YousfiEMJ2012, Kushner2012, Boeuf2013}. Studies of breakdown phenomena at atmospheric pressure in short non-uniform air gaps \cite{Naidis2005} or in microwave fields \cite{ChaudhuryBZP2011} benefit from related streamer studies. Streamers appear as well in resistive plate chambers used in high-energy physics, where the streamer mode of operation must be carefully controlled and optimized \cite{Fonte2002, KhorashadEM2011, DenniFFMPSP2011}.

The progress of streamer simulations with fluid models has recently been reviewed in~\cite{LuqueE2012}. The main point for the present paper is that essentially all these numerical studies are based on the classical fluid model, as we have called it in~\cite{LiEH2010, LiEH2012, LiTNHE2012}. The classical fluid model for the electron density contains a reaction term for impact ionization and attachment, a drift term accounting for electron displacement in the local field, and electron diffusion; the reaction term is typically taken as a function of the local electric field. The structure of this classical model with reaction, drift and diffusion is based on basic symmetry considerations and conservation laws; it has emerged as a purely phenomenological model in the course of the past century.

On the other hand, cross-sections for the collisions of electrons with molecules have become available with growing accuracy \cite{Crompton1994, PetrovicSNDSJMS2007, PetrovicDMMNSJSR2009, Raju2012, AnzaiKHTICBBCBGVI2012}. They are the input either for the Boltzmann equation or for Monte Carlo models. These microscopic particle models have not yet been appropriately linked with the much older phenomenological fluid models. This becomes evident when comparing their solutions. A negative streamer ionization front in nitrogen represented by a Monte Carlo model cannot appropriately be described by a classical fluid model even if the transport and reaction coefficients for the fluid model are derived from the Monte Carlo model \cite{LiBEM2007}. In \cite{LiBEM2007}, the coefficients for the fluid model were derived as so-called bulk coefficients; therefore in this model the electron swarms evolve correctly, and ionization fronts in a given electric field propagate with the correct velocity, as they are so-called puled fronts~\cite{LiBEM2007}. However, the electron density behind the ionization front is too low. As we will see in section~\ref{sec3} of the present paper, a classical fluid model evaluated with flux coefficients gives better ionization densities behind the front, but too low front velocities.

A phenomenological extension of the classical fluid model with a gradient expansion considerably improves the approximation of the microscopic electron dynamics~\cite{Aleksandrov1996, Naidis1997}, and it cures the deficiency of the local field approximation by including a dependence of the impact ionization rate on the local electron density gradient~\cite{LiEH2010}. That the extended fluid model matches the Monte Carlo model well up to the moment of streamer branching, can also be seen in a recent comparison of three-dimensional model simulations in~\cite{LiTNHE2012}. On the other hand, the model is still based on a local field approximation, therefore electron energies in the streamer interior erroneously are modeled as relaxing instantaneously to the low field in the streamer interior.

A full simulation by Monte Carlo models would deliver the physically most reliable results, but this is computationally extremely costly. As a compromise between accuracy and efficiency, hybrid methods \cite{LiTNHE2012, LiEH2010, LiEH2012, LiEBH2008} have been developed in the past years to track the fast and energetic processes in the ionization front with a particle model, and to model the many electrons at lower fields in the streamer interior with a fluid model. Fluid models are also the most efficient for the long time evolution in the streamer interior where slow chemical reactions and a slow thermalization of the deposited energy sets in.

These demands on fast and quantitative streamer simulations ask for the development of a fluid model that accurately incorporates the microscopic electron dynamics contained in Boltzmann or Monte Carlo models. We will here follow the strategy to derive such a model not from phenomenological considerations, but through a systematic derivation from the Boltzmann equation.

This paper is the first in an ongoing investigation of high order fluid models for streamer discharges; it is focused on the derivation of first and higher order fluid models and on the derivation and correct implementation of transport data in these models. Section \ref{sec2} describes the derivation of models of different order. The starting point is the set of balance equations obtained as velocity moments of the Boltzmann equation. Then the derivation of the classical first order model is briefly described, and previous approaches to higher order models are summarized. In section \ref{sec2d} our new high order fluid model is derived in a systematic manner. The balance/moment equations are closed after the balance equation for the energy flux. This is done by approximating the high order tensors in the energy flux balance equation by a product of two lower moments while the collision transfer terms are evaluated using momentum transfer theory. In section 3, electron transport properties as an input in fluid models are calculated using a multi term theory for solving the Boltzmann equation \cite{WhiteRDNL2009, DujkoWPR2010}. We pay particular attention to the accuracy of the calculation and the proper use in both the first and high order fluid models. The results of our multi term solution of Boltzmann's equation are compared with those obtained by the publicly available Boltzmann solver BOLSIG+ developed by Hagelaar and Pitchford~\cite{{HagelaarP2005}} for electrons in molecular nitrogen. BOLSIG+ is a popular Boltzmann solver based on a classical two term theory so we have been motivated to check its accuracy and integrity against the advanced and highly sophisticated multi term Boltzmann solver developed by the group from the James Cook University and their associates \cite{WhiteRDNL2009, DujkoWPR2010, RobsonN1986, NessR1986}. In section 4 we give the results for negative planar fronts obtained with the first order model with particular emphasis upon the consistent implementation of transport data. The results for various streamer properties obtained with different type of input data are compared. A thorough analysis of numerical streamer front solutions with our high order model, and a comparison with Monte Carlo results is contained in an accompanying second paper~\cite{PaperII}.
\section{Derivation of first and higher order fluid models}
\label{sec2}
\subsection{General considerations: Boltzmann equation and moment equations}
\label{sec2a}
Our starting point is the Boltzmann equation for charged particles in an electric field $\vc{E}$
\begin{equation}
\partial_t f_i + \vc{c}\cdot\nabla f_i + \frac{e_i}{m_i}\vc{E}\cdot\nabla_{\vc{c}} f_i = -J(f_i,f_0).
\label{2.1.1}
\end{equation}
Here $f_i(\vc{r},\vc{c},t)$ is the distribution function in phase space at the position $\vc{r}$ and velocity $\vc{c}$ for each charged component $i$, $\nabla$ is the differential operator with respect to space $\vc{r}$ and $\nabla_{\vc{c}}$ with respect to velocity $\vc{c}$, $e_i$ and $m_i$ are charge and mass of species $i$, and $t$ is time. The right-hand side of equation (\ref{2.1.1}), $J(f_i,f_0)$, describes the collisions of charged particles with neutral molecules, accounting for elastic, inelastic, and non-conservative (e.g. ionizing or attaching) collisions, and $f_0$ is the velocity distribution function of the neutral gas (usually taken to be Maxwellian at temperature $T_g$).

Streamer discharges have a characteristic nonlinear coupling between densities of charged particles and electric field. The space charge modifies the field, and the field determines the drift, diffusion and rate coefficients. The electric field has to be calculated self-consistently with Poisson's equation
\begin{equation}
\nabla\cdot\vc{E}=\frac{1}{\epsilon_0}\sum_{i}e_i\;n_i(\vc{r},t),
\label{2.1.2}
\end{equation}
where $\epsilon_0$ is the dielectric constant, and $e_i$ and $n_i$ are the charges and densities of species $i$ that can be electrons and ions. As ions are much heavier than electrons, their motion is typically neglected within the ionization front, and the analysis focuses on the electron evolution. Thus, in what follows we suppress the charged particle index $i$ in equation (\ref{2.1.1}) and focus in the electron dynamics.

After solving equations (\ref{2.1.1}) and (\ref{2.1.2}), quantities of physical interest could be obtained as velocity `moments' of the distribution functions, starting with the number density
\begin{equation}
n(\vc{r},t) = \int f(\vc{r},\vc{c},t)d\vc{c},
\label{2.1.3}
\end{equation}
followed by higher order quantities
\begin{equation}
\big\langle\phi(\vc{c})\big\rangle = \frac{1}{n(\vc{r},t)}\int \phi(\vc{c})f(\vc{r},\vc{c},t)d\vc{c},
\label{2.1.4}
\end{equation}
with $\phi(\vc{c})=m\vc{c},\frac{1}{2}mc^2,\ldots$ furnishing the average velocity $\vc{v}=\langle \vc{c} \rangle$, average energy $\varepsilon=\langle \frac{1}{2}mc^2 \rangle$, and so on.

Equations (\ref{2.1.1}) and (\ref{2.1.2}) represent a non-linear coupled system of partial differential equations in 6-dimensional phase space and in time, with a complicated collision operator $J$. For a complete problem description, appropriate initial and boundary conditions on $f(\vc{r},\vc{c},t)$ in phase space must be implemented.

For streamers, it is very difficult to solve equations (\ref{2.1.1}) and (\ref{2.1.2}) due to the large gradients of electric fields and of charged particle densities at their fronts. Moreover, the streamers create a self-consistent field enhancement at their tips which allow them to penetrate into regions where the background field is too low for efficient ionization processes to take place. From a kinetic theory point of view, this is a non-stationary, non-hydrodynamic and non-linear problem where space and time should be treated on an equal footing with velocity in equation (\ref{2.1.1}). Clearly, the numerical solution of equations (\ref{2.1.1}) and (\ref{2.1.2}) for the full streamer problem is a formidable task and a fluid equation treatment is more tractable. In a fluid approach, the problem of solving the Boltzmann equation for $f$ in phase-space is replaced by a set of low order approximate (velocity) moment equations of $f$ \cite{WhiteRDNL2009, RobsonWP2005, RobsonNLW2008}.

Fluid equations may be derived either directly as moments of (\ref{2.1.1}) or from first principles using physical arguments. Following the first approach, the set of moment/balance equations can be found by multiplying (\ref{2.1.1}) by $\phi(\vc{c})$ and integrating over all velocities
\begin{equation}
\partial_t\;\big(n\big\langle \phi(\vc{c})\big\rangle\big) + \nabla\cdot\big(n\big\langle\vc{c}\phi(\vc{c})\big\rangle\big)
-n\frac{e}{m}\vc{E}\cdot\big\langle\nabla_{\vc{c}}\phi(\vc{c})\big\rangle=C_{\phi}\,,
\label{2.1.5}
\end{equation}
where $<>$ represents the average over the velocity $\vc{c}$ of the charged particles, and $C_\phi$ is the collision term in the balance equation:
\begin{equation}
C_{\phi} = -\int \phi(\vc{c})J(f)d\vc{c}\,.
\label{2.1.6}
\end{equation}
To derive the term $\big\langle\nabla_{\vc{c}}\phi(\vc{c})\big\rangle$ with its negative sign, a partial integration over $\vc{c}$ has been performed.

If one now takes $\phi(\vc{c})$ consecutively equal to $1$, $m\vc{c}$, $\frac{1}{2}mc^2$ and $\frac{1}{2}mc^2\vc{c}$, etc., equation (\ref{2.1.5}) generates an infinite series of equations, a full solution of which would be equivalent to calculating $f$ itself. In practice, one truncates after a certain moment equation; in this process obviously some information of the Boltzmann equation is lost. Therefore we now discuss the derivation of first and second order fluid models with their truncations of the moment equations, and then we derive systematically our new high order fluid model.

\subsection{The first order or classical fluid model}
\label{sec2b}
First we derive the classical fluid model as an approximation from the Boltzmann equation. We use continuity and momentum balance equation, i.e., $\phi(\vc{c})=1$ and $m\vc{c}$, and truncate the set (\ref{2.1.5}) at the momentum balance equation,
\begin{equation}
\partial_t n + \nabla \cdot \big(n\vc{v}\big) = C_1\,,
\label{2.2.1}
\end{equation}
\begin{equation}
\partial_t(nm\vc{v}) + \nabla\cdot( nm\langle \vc{c}\vc{c}\rangle) - ne\vc{E} = C_{m\bf{c}}\,,
\label{2.2.2}
\end{equation}
where $\vc{v}=\langle\vc{c}\rangle$ is the average local electron velocity. Now the velocities are decomposed into an average velocity $\vc{v}$ plus random velocities $\vc{c}-\vc{v}$ with zero mean ($\langle\vc{c}-\vc{v}\rangle=0$). On introducing the pressure tensor
\begin{equation}
\label{pressure}
\vc{P}=nm\langle (\vc{c}-\vc{v})(\vc{c}-\vc{v}) \rangle\,,
\label{2.2.3}
\end{equation}
equation (\ref{2.2.2}) becomes
\begin{equation}
\partial_t\left( nm\vc{v}\right)+\nabla\cdot\left( nm\vc{v}\vc{v}\right)+\nabla\cdot\vc{P}-ne\vc{E} = C_{m\bf{c}}
\label{2.2.4}
\end{equation}
where the following identity was used
\begin{eqnarray}
\nabla \cdot \left( nm\langle \vc{c}\vc{c} \rangle \right) &= \nabla \cdot \big[ nm \big( \vc{v}\vc{v}+\vc{v}\langle \vc{c}-\vc{v}\rangle + \langle \vc{c}-\vc{v}\rangle\vc{v}
 +\langle \left( \vc{c}-\vc{v} \right) \left( \vc{c}-\vc{v}\right) \rangle \big)\big] \nonumber \\
 &=\nabla \cdot \big( nm \;\vc{v}\vc{v}\big) +\nabla \cdot \big(nm\;\langle \left( \vc{c}-\vc{v} \right) \left( \vc{c}-\vc{v}\right) \rangle \big)\big]
\label{2.2.5}
\end{eqnarray}
The second term in the left-hand side of (\ref{2.2.4}) can be expanded as
\begin{equation}
\nabla\cdot\left(nm\;\vc{v}\vc{v}\right) = nm\left(\vc{v}\cdot\nabla\right)\vc{v}+\vc{v}\Big[\nabla\cdot\left( nm\vc{v}\right)\Big].
\label{2.2.6}
\end{equation}
Now we substitute (\ref{2.2.6}) into (\ref{2.2.4}), use the continuity equation (\ref{2.2.1}) and introduce the convective time derivative
\begin{equation}
\frac{d}{dt} = \partial_t + \vc{v}\cdot\nabla,
\label{2.2.8}
\end{equation}
which measures the rate of change in a reference frame moving with the mean drift velocity $\vc{v}$ of the electrons. The momentum balance equation then obtains the form
\begin{equation}
nm\frac{d\vc{v}}{dt} = ne\vc{E} - \nabla\cdot\vc{P} + C_{m\bf{c}} - m\vc{v}C_1.
\label{2.2.9}
\end{equation}
The physical interpretation of this equation is straightforward: the rate of change of the mean electron velocity is due to the force of the electric field plus forces due to the pressure of the electron fluid itself and due to internal forces associated with the collisional interactions with a large number of neutral gas molecules at rest. It should be emphasized, however, that the form with the convective derivative $d/dt$ is not useful for the analysis of a full streamer problem where local fields and therefore local mean electron velocities $\vc v$ vary largely in space and time.

Because the system has been truncated, the yet unspecified tensor $\vc{P}$ (\ref{2.2.3}) of electron pressure appears on the right-hand side of (\ref{2.2.9}), and a closure assumption needs to be found. If the distribution of random velocities is close to isotropic, the diagonal terms of $\vc{P}$ are equal and given by the scalar kinetic pressure $p$,  
\begin{equation}
\vc{P}\approx p \;\vc{I}=nkT\,,
\label{2.2.10}
\end{equation}
where $\vc{I}$ is the unity tensor, $k$ is the Boltzmann constant and $T$ is the temperature. It should be emphasized here that the isotropy of the velocity fluctuations and of the pressure is a strong assumption at the streamer tip where the electric fields are high and strong pressure gradients exist. In this streamer region the random spread of electrons along the field direction can differ significantly from the perpendicular direction.

The next simplifying assumption concerns the collision terms. An expression often used for momentum transfer by collision is
\begin{equation}
C_{m\bf{c}}= -nm\nu_{\rm eff}\vc{v}\,,
\label{2.2.11}
\end{equation}
which assumes that the force per unit volume exerted on the electrons due to collisions with neutral molecules is proportional to the average electron velocity. The proportionality constant is called the effective momentum transfer collision frequency; it accounts for momentum exchange in elastic and inelastic collisions. With this simplifying assumption and neglecting the transfer of momentum in non-conservative collisions\footnote{Non-conservative collisions do not conserve particle numbers as they account for ionization, attachment or recombination reactions.} relative to other momentum transfer collisions (which is usually a good approximation), the momentum balance equation (\ref{2.2.9}) becomes
\begin{equation}
nm\frac{d\vc{v}}{dt} = ne\vc{E} - \nabla p - nm\nu_{\rm eff}\vc{v}.
\label{2.2.12}
\end{equation}
If the rate of momentum change $(d_t \vc v)/\vc v$ is smaller than the rate of momentum transfer $\nu_{\rm eff}$, and if the gradients in electron energy can be neglected, one gets the following expression for the average flux of the electrons
\begin{equation}
\vc{\Gamma}=n\vc{v}=n\mu\vc{E}-D\nabla n\,,
\label{2.2.13}
\end{equation}
where mobility and diffusion constant are given by
\begin{equation}
\mu = \frac{e}{m\nu_{\rm eff}}\,,\quad D=\frac{kT}{m\nu_{\rm eff}}\,,
\label{2.2.14}
\end{equation}
if the system is close to equilibrium. In this case the Nernst-Townsend-Einstein relation
\begin{equation}
\frac{D}{\mu}=\frac{kT}{e}
\label{2.2.15}
\end{equation}
is valid. The steady-state form of equation (\ref{2.2.12}) is an acceptable approximation because the effective time constant for momentum transfer $\frac{1}{\nu_{\rm eff}}$ at atmospheric pressure is generally much less than the time scale on which the local electric field varies within a streamer \cite{LiBEM2007, KanzariYH1998}.

Further from equilibrium as in the head of the streamer where the electric fields are high, the approximation (\ref{2.2.10}) is not valid as the velocity fluctuations of the electrons are clearly anisotropic. Let us consider once more the special case where the average velocity of electrons is stationary. From equation (\ref{2.2.9}) and neglecting momentum transfer in non-conservative collisions, we obtain
\begin{equation}
mn\nu_{\rm eff}\vc{v} = en\vc{E} - \nabla\cdot\vc{P}\,,
\label{2.2.16}
\end{equation}
and as a result for the flux we have
\begin{equation}
\vc{\Gamma} = n\mu\vc{E} - \frac{1}{\nu_{\rm eff}}\nabla\cdot\big[n\langle\left(\vc{c}-\vc{v}\right)\left(\vc{c}-\vc{v}\right)\rangle\big].
\label{2.2.17}
\end{equation}
Comparing the last equation and equation (\ref{2.2.13}), it is obvious that quantity
\begin{equation}
\vc{D}^{\star} = \frac{\langle \left(\vc{c}-\vc{v}\right)\left(\vc{c}-\vc{v}\right)\rangle}{\nu_{\rm eff}}
\label{2.2.18}
\end{equation}
is the reminisce of the diffusion tensor often used in the drift-diffusion approximation instead of the diffusion constant. However, the physical interpretation of this quantity is not {\it a priori} clear. Though this quantity assumes the anisotropic nature of the temperature tensor, it reduces to the diffusion constant $D$ when the effective momentum transfer collision frequency $\nu_{\rm eff}$ is independent of the electron energy. For an energy-dependent effective momentum transfer collision frequency, the straightforward generalization of the Einstein relation for the diffusion constant (\ref{2.2.14}) to the diffusion tensor $\vc{D}$ or application of $\vc{D}^{\star}$ in the equation for the average flux (\ref{2.2.13}) is misleading and reader is referred to \cite{Robson1984, MasonMcD1988, Robson1986} where the so-called generalized Einstein relation were introduced using the momentum and energy balance equations. However, the generalization of the diffusion constant $D$ appearing in (\ref{2.2.13}) to diffusion tensor $\vc{D}$ is a very welcome step in fluid modeling of streamer discharges due to the often strong anisotropic nature of the diffusion tensor for certain gases and due to deviations of the electron velocity distribution function from a Maxwellian distribution of velocities in different spatial regions of the streamers.

When the above approximation for the particle flux is inserted into equation (\ref{2.2.1}), we get the well known reaction drift diffusion expression for the charged particle motion in the discharge
\begin{equation}
\partial_t n + \nabla \cdot \big(\mu(\vc E)\vc{E}n - \vc D(\vc E) \cdot \nabla n\big) = C_1\,.
\label{2.2.19}
\end{equation}

Originally this approximation was not derived from the Boltzmann equation, but derived on purely phenomenological grounds. It is clear that the equation must have the structure of a continuity equation with a source term for electron generation and loss. The parametric dependence of the source term is open at this point; within the classical model the source term is assumed to depend on local particle densities and on the local electric field. The second approximation concerns the electron velocity $\vc{v}$. On a time scale much larger than the collision frequency, the electron motion is overdamped, so the electrons must drift against the direction of the electric field according to the second term in (\ref{2.2.19}). The stochastic and undirected motion is modeled in an {\it ad hoc} manner through the diffusion term; the fact that the diffusion correction is added to the drift term and not included in any other functional manner is not {\it a priori} clear and can actually be deduced from the above analysis.

In conclusion, the lowest level of fluid approximation, also called classical model or first order model, is given by the equations
\begin{eqnarray}
&\frac{\partial n}{\partial t}  = \nabla \cdot \left(\vc{D}\cdot \nabla n\right) + \nabla\cdot(\mu n\vc{E}) + n(\nu_I - \nu_A)\,, \label{2.2.20} \\
&\frac{\partial n_p}{\partial t}  = n\nu_I\,, \label{2.2.21} \\
&\frac{\partial n_n}{\partial t}  = n\nu_A\,, \label{2.2.22}
\end{eqnarray}
coupled to the Poisson equation for the electric field,
\begin{equation}
\nabla^{\,2}\phi = \frac{e}{\epsilon_0}(n-n_p-n_n), \quad \vc{E} = -\nabla\phi\,.
\label{2.2.23}
\end{equation}
Here $n_p$ and $n_n$ are positive and negative ion densities, while $\nu_I$ and $\nu_A$ are the ionization and attachment collision frequencies due to electron-molecule collisions, and $\phi$ is the electric potential. The numerical solution of the system (\ref{2.2.20})-(\ref{2.2.23}) requires the transport properties $\mu$, $\vc{D}$, $\nu_I$ and $\nu_A$ as a function of the local electric field for the gas in question. The derivation and implementation of transport data is discussed in section~3. The numerical solutions of planar streamer fronts with these transport data are discussed in section 4.
\subsection{Second order models including the energy balance equation}
\label{sec2c}

We now turn to models that include the second velocity moment of the Boltzmann equation, i.e., the energy balance equation. We will see that the closure of this equation is not a straight forward process, and we will discuss some approximations made in the literature. The results of a second order model and of our high order model for planar streamer fronts will be compared in our second paper~\cite{PaperII}.

The first important steps beyond first order fluid models of streamer discharges, to our knowledge, were carried out by Abbas and Bayle \cite{AbbasB1980, AbbasB1981}, and by Bayle and Carenbois \cite{BayleC1985}. They employed a second order model which involves the energy balance equation to explore the zone at the streamer tip where the electron energy is not determined anymore by the local electric field only. However, as pointed out by Kanzari {\it et al.}~\cite{KanzariYH1998}, the accuracy of their model was restricted by the drastic assumptions taken in their energy balance equation. They evaluated the source term in the energy balance equation and corresponding averages by assuming a Maxwellian distribution for the electrons. Guo and Wu \cite{GuoW1993} have developed a more sophisticated second order model in which the Langevin theory was used to simplify the collision source terms with {\it a priori} knowledge of the relaxation times of electron energy and momentum. Kanzari {\it et al.} \cite{KanzariYH1998} have made an important step further by calculating the source term in the energy equation in a more consistent way where the {\it ad hoc} assumptions for the distribution function were avoided. A similar approach was later used by Eichwald {\it et al.} \cite{EichwaldDMYD2006} in their simulations of the streamer dynamics and of the radical formation in a pulsed corona discharge used for flue gas control. Their results showed that the average electron velocity in the streamer head is largely overestimated by the classical first order model. As a consequence, electron density and radical density in the ionized channel were up to 50\% higher than with the second order model. The salient feature of their theory is the fact that the heat flux term in the energy balance equation was explicitly neglected, but this, unfortunately, is of questionable accuracy. We will illustrate in streamer simulations in our second paper~\cite{PaperII} that the energy flux plays an important role in determining the streamer behavior. Thus, particular care should be taken with the closure through specification of the energy flux. Even for charged particle swarms under non-hydrodynamic conditions one must be careful how to specify the heat flux vector \cite{WhiteRDNL2009, RobsonWP2005, RobsonNLW2008, NicoletopoulosR2008}. We now discuss how the fluid equations should be closed for streamers, while we stress that the method itself is applicable to a much wider range of phenomena.
\subsection{High order fluid model}
\label{sec2d}

We now derive a fluid model including the energy flux equation, i.e., we truncate at the moment equation with the third power of velocity. We will argue that the energy flux equation is crucial for the success of the fluid model for streamers. Inserting different moments of $\vc{c}$ into equation (\ref{2.1.5}), one finds
\begin{equation}
\frac{\partial n}{\partial t} + \nabla \cdot n\vc{v} = C_1\,,
\label{2.3.1}
\end{equation}
\begin{equation}
\frac{\partial}{\partial t}\left(nm\vc{v}\right) + \nabla\cdot\left( nm\langle \vc{c}\vc{c}\rangle\right) - ne\vc{E} = C_{m\bf{c}}\,,
\label{2.3.2}
\end{equation}
\begin{equation}
\frac{\partial}{\partial t}\left(n\Big\langle \frac{1}{2}mc^2 \Big\rangle \right) + \nabla\cdot\left( n \Big\langle \frac{1}{2}mc^2\vc{c} \Big\rangle \right) -ne\vc{E}\cdot\vc{v} = C_{\frac{1}{2}mc^2}\,,
\label{2.3.3}
\end{equation}
\begin{equation}
\frac{\partial}{\partial t}\left(n\Big\langle \frac{1}{2}mc^2\vc{c} \Big\rangle \right) +  \nabla\cdot\left( n \Big\langle \frac{1}{2}mc^2\vc{c}\vc{c} \Big\rangle \right)-ne\vc{E}\cdot\Bigg\langle \frac{\partial}{\partial \vc{c}}\Big(\frac{1}{2}mc^2\vc{c} \Big)\Bigg\rangle = C_{\frac{1}{2}mc^2\bf{c}}\,.
\label{2.3.4}
\end{equation}
Different forms of these equations that some readers might be familiar with, are given in the appendix. It will be shown in section \ref{sec2g} that the explicit form of the high order tensors appearing in the divergence term of the energy flux equation (\ref{2.3.4}) is not required.

This system of equations is exact at this stage, but not very useful and the reason is twofold. First, there are more unknowns than equations, the familiar problem of closure. The set of fluid equations can be increased to an arbitrary size merely by taking higher velocity moments of the Boltzmann equation. Second, the definition of the system requires the respective collision terms $C_\phi$.

\subsubsection{Collision processes.}
\label{sec2e}
In contrast to Monte Carlo simulations or kinetic equations in which cross sections for charged particle scattering enter into the calculations in a fairly clear way, in fluid equations the collisions are treated in a variety of ways which are not necessarily consistent with the data itself and/or with the system under consideration. There are many examples which illustrate this issue and the reader is referred to \cite{WhiteRDNL2009,RobsonWP2005,RobsonNLW2008} for a detailed discussion. In the context of streamer and breakdown studies beyond first order fluid models, it has become common practice to evaluate the collision terms and averages by assuming some particular form of the velocity distribution function, usually Maxwellian \cite{BayleC1985,GuoW1993,SobotaMVDH1010,DavoudabadiSM2009, SimaPYYS2012}.

Let us consider the form of the collision terms in the balance equations (\ref{2.3.1})-(\ref{2.3.4}). In this work we characterize the elastic, inelastic and non-conservative collisions by corresponding average collision frequencies. A collision frequency $\nu(v_r)$ for collisions between charged particles and gas molecules is related to the cross section $\sigma(v_r)$ characterizing the process by
\begin{equation}
\nu(v_r) = n_gv_r\sigma(v_r)\,,
\label{2.4.1.1}
\end{equation}
where $\vc{v}_r$ is the relative velocity and $n_g$ is the number density of the neutral gas. Further, we deal with weakly ionized systems where the interactions of charged particles with one another and with excited molecules is negligible.

Our calculation takes into account that the charged particles can loose energy and momentum even in elastic collisions with the gas molecules as the finite mass and the thermal energy of the molecules are taken into account. The momentum exchange in inelastic collisions is included and $\nu_m=\nu_m(v_r)$ denotes the total momentum transfer collision frequency while $\nu_i$ and $\nu_i^{\;s}$ are inelastic and superelastic collision frequencies for the inelastic channel $i$. The total collision frequencies for attachment and ionization are denoted by $\nu_A=\nu_A(v_r)$ and $\nu_I=\nu_I(v_r)$, respectively. We consider only single ionization with ionization energy $\epsilon_I$, but the resulting ion can be left in any one of its internal excited states, characterized by an excitation energy $\Delta\epsilon_I^{(i)}$ and a collision frequency $\nu_I^{(i)}$.

\subsubsection{Momentum transfer theory.}
\label{sec2f}
The momentum transfer theory has a long history in kinetic theory of gases and has proved very successful for describing charged particle transport in gases under the influence of electric and magnetic fields. It is discussed comprehensively in the textbook of Mason and McDaniel \cite{MasonMcD1988} and in references \cite{WhiteRDNL2009, RobsonWP2005, RobsonNLW2008, Robson1986, RobsonN1988, VrhovacP1996, LiWR2001}. The main ideas can be briefly summarized as follows:

\noindent(1) We need to express the average collision frequencies as a function of the mean energy in the center-of-mass reference frame. Thus, we replace the variables
\begin{equation}
v_r\rightarrow\varepsilon=\frac{1}{2}\mu_r v_r^2\,,
\label{2.4.2.1}
\end{equation}
in expressions for collisional frequencies
\begin{equation}
\nu=\nu(v_r)\rightarrow\widetilde{\nu}=\widetilde{\nu}(\varepsilon)\,,
\label{2.4.2.2}
\end{equation}
where $\varepsilon$ is the energy measured in the center-of-mass reference frame and $\mu_r$ is the reduced mass.

\noindent(2) If we assume that the dominant contributions to the averages come from regions near the mean energy $\bar{\varepsilon}$, and that $\nu(\varepsilon)$ varies sufficiently slowly with $\varepsilon$, then a Taylor expansion
\begin{equation}
\widetilde{\nu}(\varepsilon) = \widetilde{\nu}(\bar{\varepsilon})+\widetilde{\nu}^{\;'}(\bar{\varepsilon})(\varepsilon-\bar{\varepsilon})+\ldots\,,
\label{2.4.2.3}
\end{equation}
is expected to be a reasonable approximation. For conservative collisional processes such as elastic and inelastic scattering, only the first term of the expansion (\ref{2.4.2.3}) is considered. However, when energy-dependent non-conservative processes such as ionization and electron attachment are operative then the derivative term in (\ref{2.4.2.3}) becomes the leading term and must be included.

\noindent(3) Using the momentum transfer approximation, the balance equations (\ref{2.3.1})-(\ref{2.3.4}) get the form
\begin{equation}
\frac{\partial n}{\partial t} + \nabla \cdot n\vc{v} = -n\big(\widetilde{\nu}_A - \widetilde{\nu}_I\big)\,,
\label{2.4.2.4}
\end{equation}
\begin{equation}
\frac{\partial}{\partial t}(mn\vc{v})+\nabla\cdot\big( mn\langle \vc{c}\vc{c}\rangle \big)-ne\vc{E}=-\mu_r n\widetilde{\nu}_m\vc{v}-mn\vc{v}\big[ \widetilde{\nu}_I+\zeta\widetilde{\nu}_A^{\,'}\big]\;,
\label{2.4.2.5}
\end{equation}
\begin{eqnarray}
\frac{\partial}{\partial t}(n\varepsilon)+&\nabla\cdot(n\vc{\xi})-ne\vc{E}\cdot\vc{v} = -n\widetilde{\nu}_e\left(\varepsilon-\frac{3}{2}kT_0 \right)\nonumber \\
&-n\frac{m_0}{m+m_0}\sum_{i}(\widetilde{\nu}_i-\widetilde{\nu}_i^{\;s})\epsilon_i-
n\varepsilon\widetilde{\nu}_A-n\sum_i\widetilde{\nu}_i^{(i)}\Delta\epsilon_i^{(i)}\;,
\label{2.4.2.6}
\end{eqnarray}
\begin{equation}
\frac{\partial}{\partial t}(n\vc{\xi}) + \nabla\cdot\left( n \Big\langle \frac{1}{2}mc^2\vc{c}\vc{c} \Big\rangle \right)-ne\vc{E}\cdot\Bigg\langle \frac{\partial}{\partial \vc{c}}\Big(\frac{1}{2}mc^2\vc{c} \Big)\Bigg\rangle = -n\widetilde{\nu}_m\vc{\xi}\,,
\label{2.4.2.7}
\end{equation}
where $\varepsilon$ is the average electron energy, $\vc{\xi}$ is the average electron energy flux and $m_0$ is the mass of gas molecules. $\zeta$ is given by \cite{Robson1986, VrhovacP1996}:
\begin{equation}
\zeta = \frac{2}{3}\frac{m_0}{m+m_0}\left(\frac{1}{2}m\big\langle c^2 \big\rangle-\frac{1}{2}m|\vc{v}|^2\right)\,.
\label{2.4.2.8}
\end{equation}
The average collision frequencies for momentum and energy transfer
\begin{equation}
\widetilde{\nu}_m(\bar{\varepsilon})=n_g\sqrt{\frac{2\varepsilon}{\mu_r}}\sigma_m(\bar{\varepsilon})\,,
\label{2.4.2.9}
\end{equation}
\begin{equation}
\widetilde{\nu}_e(\bar{\varepsilon})=\frac{2\mu_r}{m+m_0}\widetilde{\nu}_m(\bar{\varepsilon})\,,
\label{2.4.2.10}
\end{equation}
are prescribed functions of the mean energy in the centre-off-mass frame
\begin{equation}
\bar{\varepsilon}=\frac{m_0\varepsilon+m\frac{3}{2}kT_0}{m+m_0}\,,
\label{2.4.2.11}
\end{equation}
where $k$ is Boltzmann's constant and $T_0$ is the neutral gas temperature. Other collision frequencies describing the inelastic, non-conservative and superelastic collisions appearing in equations (\ref{2.4.2.4})-(\ref{2.4.2.7}) are also functions of $\bar{\varepsilon}$. It should be emphasized that the equations of continuity (\ref{2.4.2.4}), of momentum balance (\ref{2.4.2.5}) and of energy balance (\ref{2.4.2.6}) are valid for charged particles of arbitrary mass while the energy flux equation (\ref{2.4.2.7}) is obtained in the approximation of $m/m_0\ll1$. High-order corrections in momentum transfer theory corrections (e.g., high-order terms in the Taylor expansion (\ref{2.4.2.3})) could be added on the right-hand side of the system (\ref{2.4.2.4})-(\ref{2.4.2.7}) if desired, without in any way changing the generality of the physical arguments associated with the closure assumptions presented below.

\subsubsection{Solution regimes and closure assumptions.}
\label{sec2g}
The closure of the system of equations (\ref{2.4.2.4})-(\ref{2.4.2.7}) requires approximations or assumptions on the form of the pressure tensor. The standard approximation for light particles such as electrons is that the pressure tensor can be taken as a scalar at the fluid level of approximation \cite{WhiteRDNL2009, RobsonWP2005, RobsonNLW2008, Robson1986}. This means that, $\bar{\varepsilon}\approx\varepsilon\gg\frac{1}{2}mv^2$, and the pressure tensor simplifies to
\begin{equation}
\vc{P}=nk\vc{T}\approx\frac{2}{3}n\varepsilon\vc{I}\,,
\label{2.4.3.1}
\end{equation}
where $\vc{T}$ is the so-called temperature tensor that characterizes energy fluctuations even if the system is not in thermal equilibrium. This form of the pressure tensor was employed in all previous swarm oriented studies \cite{WhiteRDNL2009, RobsonWP2005, RobsonNLW2008, Robson1986, NicoletopoulosR2008, RobsonN1988, VrhovacP1996}, as well as in the recent fluid models of streamer discharges \cite{KanzariYH1998,EichwaldDMYD2006}. However, if the velocity distribution significantly deviates from isotropy in velocity space, then this approximation is problematic. For ions, the distribution function in velocity space is always anisotropic (even if elastic collisions between ions and molecules are predominant) and hence any assumption on an isotropic pressure tensor is wrong for ions. This problem can be avoided by considering the temperature tensor balance equation but this in turn contains further unknowns. The reader is referred to \cite{LiWR2001} for how to treat charged particle swarms under spatially-homogeneous hydrodynamic conditions. In streamer studies, the ions are usually considered as immobile or they are modeled by a reaction-drift-diffusion and local field approximation which is a reasonable approximation on the time scale on which a streamer ionization front passes a given point in space.

The next step in the closure of the system of equations (\ref{2.4.2.4})-(\ref{2.4.2.7}) concerns the energy flux balance equation (\ref{2.4.2.7}). The third term can be simplified as follows:
\begin{eqnarray}
\frac{\partial}{\partial\vc{c}}\Big(\frac{1}{2}mc^2\vc{c} \Big)
&=mc\frac{\partial c}{\partial\vc{c}}\;\vc{c}+\frac{1}{2}mc^2\frac{\partial\vc{c}}{\partial\vc{c}}
= m\vc{c}\vc{c}+\frac{1}{2}mc^2\vc{I},
\label{2.4.3.2}
\end{eqnarray}
where $\vc{I}$ is the unity tensor. Assuming again as in (\ref{2.4.3.1}) that the temperature tensor is isotropic, and hence that
\begin{equation} \label{cc}
\langle \vc {c c} \rangle \approx \frac{\langle c^2 \rangle}3  \;\vc I,
\end{equation}
and after some algebra, the energy flux equation can be written as
\begin{equation}
\frac{\partial}{\partial t}(n\vc{\xi}) + \nabla\cdot\left( n \Big\langle \frac{1}{2}mc^2\vc{c}\vc{c} \Big\rangle \right)-\frac{e}{m}\vc{E}\Bigg(\frac{5}{3}n\varepsilon \Bigg) = -n\widetilde{\nu}_m\vc{\xi}\,.
\label{2.4.3.3}
\end{equation}

The second term in the energy flux balance equation (\ref{2.4.3.3}) or (\ref{2.4.2.7}) is the divergence of the fourth power of the velocity averaged over the velocity distribution, $\langle c^2\vc{c}\vc{c} \rangle$, while all other terms in the equation contain only the third power of velocity. Therefore this term, that we will call the quartic tensor, requires either the next equation in the sequence of moment equations or a closure approximation. This closure assumption must be physically transparent and consistent with the general structure of the equations. One way is to approximate the relevant term by a product of lower moments as
\begin{equation}
\Big\langle c^2\vc{c}\vc{c} \Big\rangle
\approx\beta\Big\langle c^2\Big\rangle\langle\vc{c}\vc{c}\rangle 
\approx\beta\frac{4}{3m^2}\varepsilon^2\;\vc{I}\,,
\label{2.4.3.4}
\end{equation}
where we used (\ref{cc}) for the second equality. $\beta$ is a parametrization factor, generically close to unity, when the higher order correlation term $\langle c^2\vc{c}\vc{c}\rangle- \langle c^2\rangle \langle\vc{c}\vc{c}\rangle$ can be neglected.

With these closure assumptions the system of fluid equations (\ref{2.4.2.4})-(\ref{2.4.2.7}) for the electrons becomes:
\begin{equation}
\frac{\partial n}{\partial t} + \nabla \cdot n\vc{v} = -n\big(\widetilde{\nu}_A - \widetilde{\nu}_I\big)\,,
\label{2.4.3.5}
\end{equation}
\begin{equation}
\frac{\partial}{\partial t} (n\vc{v}) + \frac{2}{3m}\nabla(n\varepsilon)-n\frac{e}{m}\vc{E} = -n\vc{v}\Big(\widetilde{\nu}_m + \widetilde{\nu}_I + \frac{2}{3m}\varepsilon\widetilde{\nu}_A^{\,'}\Big)\;,
\label{2.4.3.6}
\end{equation}
\begin{eqnarray}
\frac{\partial}{\partial t}(n\varepsilon)+&\nabla\cdot(n\vc{\xi})-e\vc{E}\cdot(n\vc{v}) = -n\widetilde{\nu}_e\left(\varepsilon-\frac{3}{2}kT_0 \right)\nonumber \\
&-n\sum_{i}(\widetilde{\nu}_i-\widetilde{\nu}_i^{\;s})\epsilon_i-
n\varepsilon\widetilde{\nu}_A-n\sum_i\widetilde{\nu}_i^{(i)}\Delta\epsilon_i^{(i)}\;,
\label{2.4.3.7}
\end{eqnarray}
\begin{equation}
\frac{\partial}{\partial t}(n\vc{\xi}) + \nabla\Big(\beta\frac{2n}{3m}\varepsilon^2 \Big)-\frac{5}{3}n\varepsilon e\vc{E} = -\widetilde{\nu}_m(n\vc{\xi})\,.
\label{2.4.3.8}
\end{equation}
The parameter $\beta$ that appears in the energy flux equation is a quantity close to unity, but can be used to fit the neglected higher order correlations. Variations of this parameter will be discussed in our second paper~\cite{PaperII}.

The system of equations (\ref{2.4.3.5})-(\ref{2.4.3.8}) has the following properties: (i) If the parameter $\beta$ is specified and if the collision terms are given then the system of equations contains no further unknowns can be numerically solved for the density, average velocity, average energy and energy flux. (ii) The equations contain mean-energy collisional rates that should be carefully derived and implemented as elaborated in the next section. (iii) Attachment enters the momentum and energy balance equation in terms of the derivative of the attachment collision frequency while for ionization only the ionization collision frequency is present. (iv) The above equations are set for a single component gas; the generalization to gas mixtures proceeds through the generalization of the collision term to a sum of terms appropriately weighted according to the mole fractions of the respective species.
\section{Transport and reaction data}
\label{sec3}
\subsection{Evaluation of the transport data: Boltzmann equation analysis}
\label{sec2h}

Here we discuss how to evaluate and implement electron transport properties in both first and high order fluid models. In plasma modeling, transport coefficients of electrons and/or ions may be obtained either from swarm experiments, from solutions of the Boltzmann equation, or from Monte Carlo simulations. We note that swarm transport data are usually given in the literature in the form of tables as a function of reduced electric field, $E/n_0$ \cite{WhiteRDNL2009, RobsonWP2005}. In fluid plasma modeling, however, several important issues have to be considered before directly implementing data from the literature.

First, transport coefficients are defined far from boundaries, sources and sinks of charged particles where the so-called hydrodynamic conditions prevail \cite{WhiteRDNL2009, DujkoWPR2010, RobsonN1986, NessR1986, DujkoEWP2011, DujkoWRP2012}. Under hydrodynamic conditions, the space-time dependence of the distribution function is expressible in terms of linear functionals of $n(\vc{r},t)$. A sufficient functional relationship between the distribution function $f(\vc{r},\vc{c},t)$ and $n(\vc{r},t)$ in the case of weak gradients is the well-known expansion
\begin{equation}
f(\vc{r},\vc{c},t) = \sum_{s=0}^{\infty}f^{(s)}(\vc{c})\odot(-\nabla)^s n(\vc{r},t)\,,
\label{2.1.43}
\end{equation}
where $f^{(s)}(\vc{c})$ are tensors of rank $s$ and $\odot$ denotes an $s$-fold scalar product. Direct application of transport properties measured/calculated in different experimental arrangements where often non-hydrodynamic conditions are present explicitly or implicitly, is problematic and should be avoided. Typical example of swarm data obtained under non-hydrodynamic conditions are those measured/calculated under the steady-state Townsend (SST) conditions \cite{SakaiTS1977,Robson1995}. Before direct application of SST data one must perform careful swarm analysis and convert the SST data into the hydrodynamic transport coefficients. Details of this procedure are presented in \cite{DujkoWP2008}.

Second, care must be taken when non-conservative collisions are significant. In the presence of non-conservative collisions there are two sets of transport coefficients, the bulk and the flux \cite{RobsonN1986, Robson1991}. Assuming the functional relationship (\ref{2.1.43}) the flux $\vc{\Gamma}$ and source term $C_1$ (usually denoted by $S(\vc{r},t)$ in previous swarm oriented studies \cite{WhiteRDNL2009, DujkoWPR2010,DujkoEWP2011, DujkoWRP2012}) appearing in the equation of continuity (\ref{2.2.1}) can  be expanded
\begin{eqnarray}
\vc{\Gamma}(\vc{r},t) &= n\vc{W}^{(\star)} - \vc{D}^{(\star)}\cdot\nabla n \label{2.1.43a} \\
S(\vc{r},t) &= S^{(0)}n(\vc{r},t) - \vc{S}^{(1)}\cdot\nabla n(\vc{r},t) + \vc{S}^{(2)}:\nabla\nabla n(\vc{r},t), \label{2.1.43b}
\end{eqnarray}
where $\vc{W}^{(\star)}$ and $\vc{D}^{(\star)}$ define, respectively, the flux drift velocity and flux diffusion tensor while $\vc{S}^{(k)}$ are expansion coefficients of the source term. Substitution of expansions (\ref{2.1.43a}) and (\ref{2.1.43b}) into the continuity equation (\ref{2.2.1}) yields the diffusion equation
\begin{equation}
\frac{\partial n}{\partial t} + \vc{W}\cdot n - \vc{D}:\nabla\nabla n = - R_an
\label{2.1.43c}
\end{equation}
which defines the bulk transport coefficients
\begin{eqnarray}
R_a &= -S^{(0)},                           \label{2.1.43d} \\
\vc{W} &= \vc{W}^{(\star)} + \vc{S}^{(1)}, \label{2.1.43e} \\
\vc{D} &= \vc{D}^{(\star)} - \vc{S}^{(2)}, \label{2.1.43f}
\end{eqnarray}
where $R_a$ is the loss rate, $\vc{W}$ is the bulk drift velocity and $\vc{D}$ is the bulk diffusion tensor.

The basic difference between the bulk and flux transport coefficients should now be apparent. The bulk drift velocity is displacement of the mean position of the electron swarm and
it characterizes the motion of the total ensemble of electrons. The presence of the electric field results in a spatial variation in the energy throughout the swarm. Under such conditions, the presence of non-conservative collisions (ionization/ attachment) may lead to a change in the position of the center-of-mass of the swarm. This effect on the bulk drift
velocity is denoted by $\vc{S}^{(1)}$. On the other hand, the flux drift velocity $\vc{W}^{(\star)}$ represents the rate of change of the position of the center-of-mass due to the electric field only and can be interpreted as the mean velocity of the electrons. Likewise the flux diffusion tensor $\vc{D}^{(\star)}$ represents the rate of spreading of the swarm due to the electric field $\vc{E}$ and gradients in density $\nabla n$. The presence of non-conservative collisions may result in the variation of $\nabla n$ throughout the swarm and a subsequent variation in the rate of change of the mean squared width of the swarm. Such effects are expressed by the second rank tensor $\vc{S}^{(2)}$. The most appropriate procedure in plasma modeling would be to use the experimental swarm data (e.g. bulk values) for the analysis of the validity of the cross section and then to calculate the flux quantities which are necessary as input data in fluid modeling. More about duality of the hydrodynamic transport coefficients and their implementation in fluid modeling can be found in the references \cite{PetrovicDMMNSJSR2009,WhiteRDNL2009, DujkoWPR2010, RobsonWP2005, DujkoEWP2011, DujkoWRP2012}.

Third, while the first order fluid model requires electron transport data as a function of local reduced electric field, what really appears in high order fluid models are mean-energy-dependent collisional rates. Since momentum transfer theory is used to determine the collision terms in the fluid equations, the most appropriate procedure would be to systematically reduce these fluid equations down to the swarm limit assuming the hydrodynamic regime. From this set of equations, one can then find relationships between collisional transfer rates and the mean energy, in a self-consistent manner. This method was employed in previous works of Robson and co-workers (see for example \cite{RobsonWP2005} and references therein), but we apply here a slightly different approach. Instead of using the so-called generalized Einstein relation to determine the mean energy from the transverse diffusion coefficient \cite{RobsonWP2005,Robson1984}, the mean energy is directly calculated from the multi term solution of Boltzmann's equation. The correspondence between the mean energy and $E/n_0$ is then used to find the correspondence between the mean energy and other relevant transport data. The momentum transfer collision rate is obtained from
\begin{equation}
\nu_m = \frac{e}{m\mu(\varepsilon)}\,,
\label{2.1.41b}
\end{equation}
where $\mu(\varepsilon)$ is the electron mobility which is here a function of the mean energy. The energy-transfer collision frequency is calculated from equation (\ref{2.4.2.10}). Thus our procedure of determining the electron transport data as an input for high order fluid models is entirely consistent with the work of Boeuf and Pitchford \cite{BoeufP1995}.

The electron transport data employed in this work are calculated using a multi term theory for solving the Boltzmann equation. The methods and techniques are by now standard and the reader is referred to our previous works \cite{WhiteRDNL2009, DujkoWPR2010, DujkoEWP2011}. Among many important aspects, we highlight the following important steps:
\begin{itemize}
\item {No assumptions on symmetries in velocity space are made, and the directional dependence of the phase-space distribution function in velocity space is represented in terms of a spherical harmonic expansion:
\begin{equation}
f(\vc{r},\vc{c},t) = \sum_{l=0}^{\infty}\sum_{m=-l}^{l}f(\vc{r},c,t)Y_m^{[l]}(\hat{\vc{c}})\,,
\label{2.1.42}
\end{equation}
where $Y_m^{[l]}(\hat{\vc{c}})$ are spherical harmonics and $\hat{\vc{c}}$ represents the angles of $\vc{c}$.
In contrast to the classical two term theory (where the sum over $l$ is performed only up to $l=1$), the number of spherical harmonics is not restricted, and our method therefore is a truly multi term approach. The differences between two-term approximation and our full multi term approach for electron transport in nitrogen are illustrated in Section \ref{sec3}. The inadequacies of a Legendre polynomial expansion (when density gradients are not parallel to the field) are highlighted in our previous publications \cite{WhiteRDNL2009, DujkoWPR2010} and avoided in this work.}
\item {As discussed above, under hydrodynamic conditions a sufficient representation of the space dependence is an expansion in terms of powers of the density gradient operator.}
\item {The velocity (energy) dependence of the phase-space distribution function is represented by an expansion about a Maxwellian at an arbitrary temperature in terms of Sonine polynomials.}
\end{itemize}

Using the appropriate orthogonality relations for the spherical harmonics and for the modified Sonine polynomials, the Boltzmann equation is converted into a hierarchy of coupled equations for the moments of the distribution function. These equations are solved numerically and all transport and rate coefficients are expressed in terms of moments of the distribution function \cite{WhiteRDNL2009, DujkoWPR2010}.
\subsection{Cross sections and transport data}
\label{sec3a}

In this section, transport and reaction coefficients for electrons in N$_2$ at a temperature of 298 K are calculated as an input for first and high order fluid models. The first order model is based on the local field approximation; it requires mobility, diffusion coefficient and ionization rate as a function of the reduced electric field $E/n_0$ (where $n_0$ is the gas number density). Compared to our previous work \cite{DujkoEWP2011}, we extend the electric field range up to 3000 Td (1 Td=$10^{-21}$ Vm$^2$). The high order fluid model requires average collision frequencies for momentum and energy transfer in elastic and inelastic collisions, and rate coefficients for all collision processes as a function of the mean electron energy.
 
We use the cross sections for electron scattering in N$_2$ provided by Stojanovi\'{c} and Petrovi\'{c} \cite{StojanovicP1998}. For elastic collision processes we use the original Boltzmann collision operator \cite{Boltzmann1872} while for inelastic processes we employ the generalization of Wang-Chang {\it et al.}~\cite{WangC1964}. The ionization collision operator is detailed in \cite{NessR1986}. We assume that the ratio between the energy of the scattered electron and the total available energy in an ionizing collision is equally distributed between 0 and 1; the same holds then, of course, for the ejected electron. We remark that at the high electron energies in the streamer tip, the assumptions on the energy division can considerably influence transport profiles. Furthermore, scattering is assumed to be isotropic. This can be problematic for high values of $E/n_0$ (generally for $E/n_0\geq1000$ Td for electrons in N$_2$) when electrons scatter predominantly in the forward direction \cite{LiEH2012,PhelpsP1985}. However, the errors in the calculated transport coefficients and rate coefficients are acceptable in fluid modeling for streamers in the range of the reduced electric fields $E/n_0$ considered in this work after appropriate renormalization of cross-sections for the scattering angle distribution \cite{LiEH2012}.

We present both {\it bulk} and {\it flux} coefficients obtained by our multi term solution of the Boltzmann equation, and we compare them with the coefficients obtained by the public available Boltzmann solver BOLSIG+ derived from the same cross sections. BOLSIG+ is based on the two term approximation \cite{HagelaarP2005} and provides exclusively flux transport data.

\subsection{Input data for the first order model}
\label{sec3c}

Figure 1a shows the electron mobility as a function of the reduced electric field $E/n_0$. We observe that bulk and flux quantities start to differ visibly above a reduced field of approximately 150 Td, this means that the ionization processes start to be significant at this value of the field. As $E/n_0$ increases further, the effect becomes more pronounced, until the bulk mobility exceeds the flux mobility by approximately 30\% at 3000 Td. The difference between bulk and flux mobility is the consequence of the spatial variation of the average electron energy within an electron swarm \cite{LiBEM2007, DujkoWPR2010, NessR1986}. If the ionization rate is an increasing function of electron energy (as is the case for the parameters considered here), electrons are preferentially created in regions of higher energy resulting in a shift in the centre of mass position as well as in a modification of the spread about the centre of mass. In nitrogen up to 3000 Td, the electrons are preferentially created at the leading edge of an electron swarm and hence the bulk mobility is larger than the flux mobility.

\begin{figure} [!htb]
a)\includegraphics[scale=0.3]{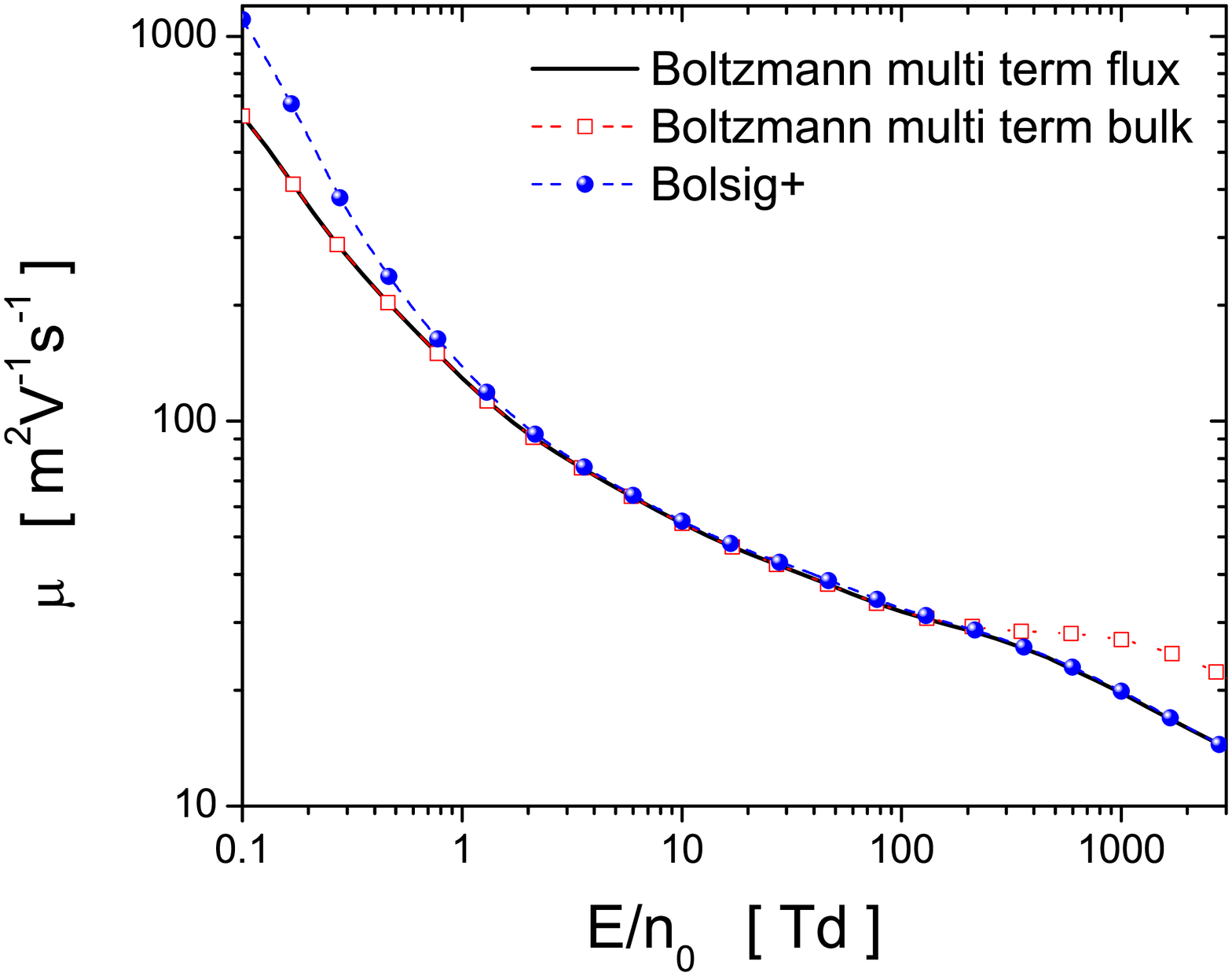}
\label{Fig1}
b)\includegraphics[scale=0.3]{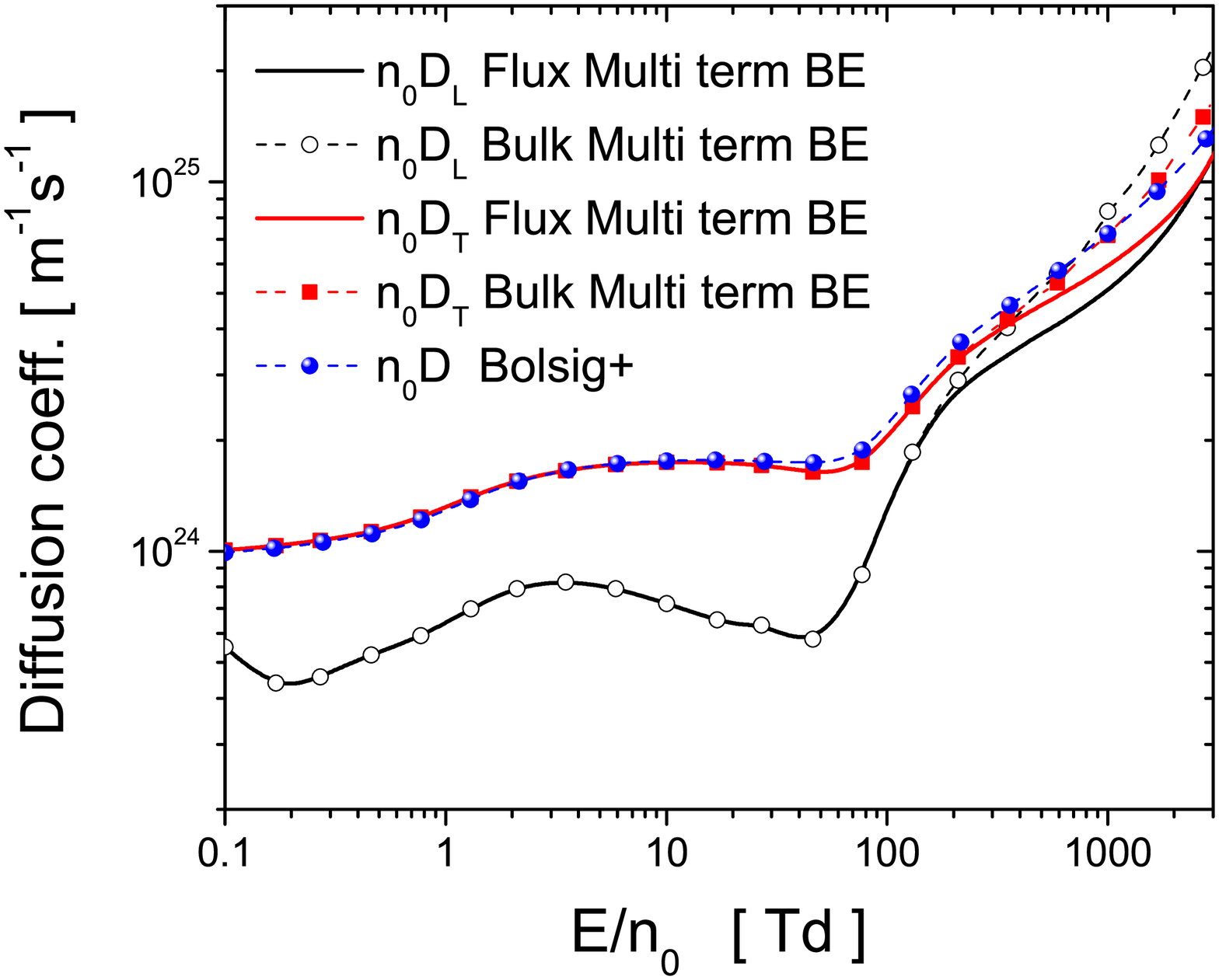}
c)\includegraphics[scale=0.3]{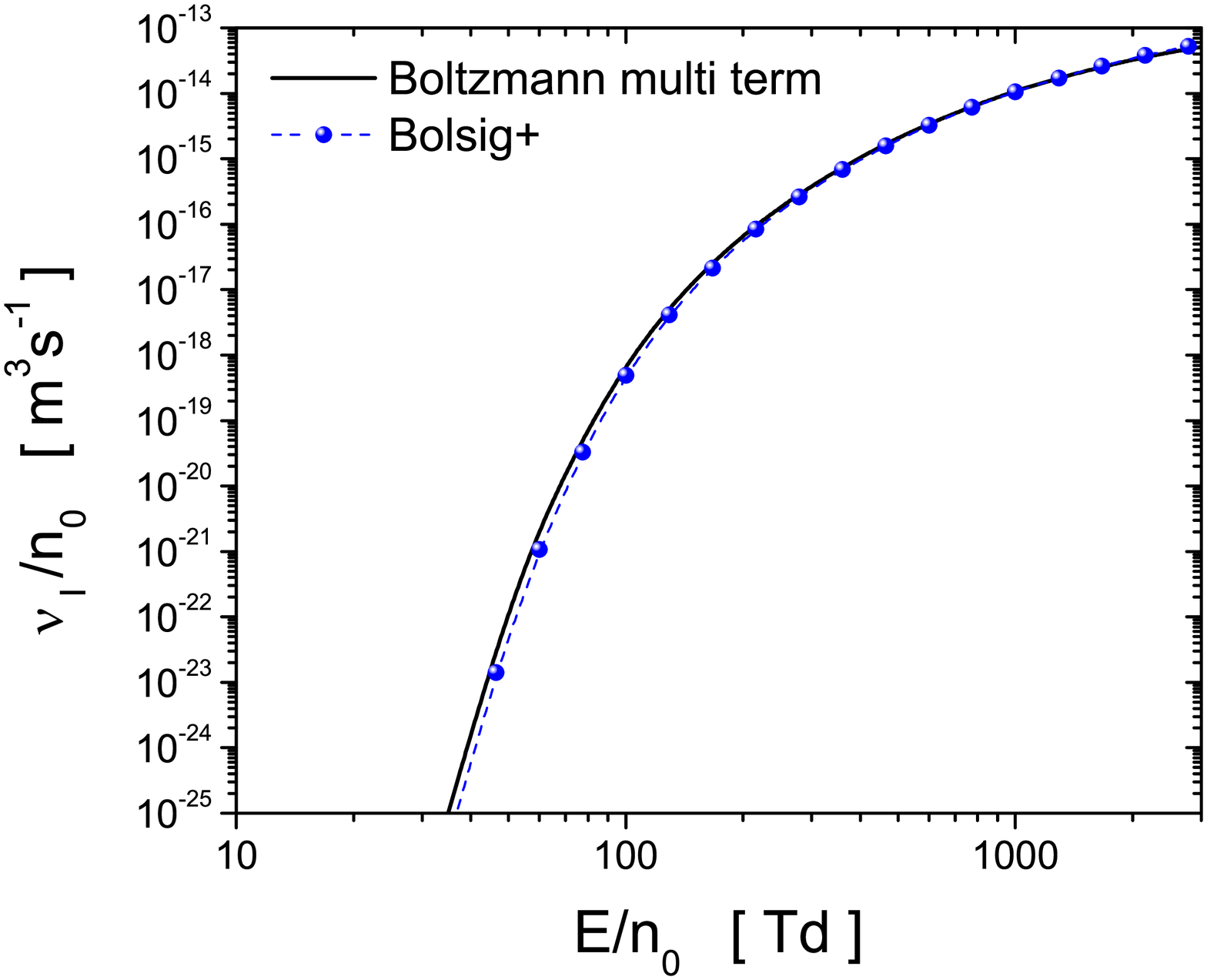}
\caption{(a) Mobility, (b) longitudinal and transverse diffusion coefficient, and (c) ionization rate of electrons in N$_2$ as a function of the reduced electric field $E/n_0$, as an input for the first order model. Shown are flux and bulk data obtained by our multi term solution of the Boltzmann equation, and flux data obtained by the BOLSIG+ code. BOLSIG+ provides only an isotropic diffusion coefficient for panel (b).}
\end{figure}

Figure 1a shows as well that the flux data obtained by our multi term solution of Boltzmann's equation and by the BOLSIG+ code agree well. Only for $E/n_0$ below about 3 Td, the BOLSIG+ mobility is higher than our flux mobility. As we have successfully compared our multi term results with Monte Carlo results that include the thermal energy of the background molecules as well, the results shown in Fig. 1a suggest that the BOLSIG+ code should be carefully tested in the limit of thermal energies.

Figure 1b shows the diffusion coefficients as a function of the reduced field $E/n_0$. The bulk and flux values of the longitudinal and transverse diffusion coefficients are compared with the isotropic diffusion coefficient calculated by the BOLSIG+ code. As for the mobility, flux and bulk data start to deviate for $E/n_0$ above approximately 150 Td, indicating again the onset of ionization effects. The diffusion coefficients are more sensitive to the ionization processes than the mobility; the differences between bulk and flux data can reach almost 50\% for $E/n_0$ approaching 3000 Td. For a more thorough analysis of the explicit influence of ionization processes on the diffusion coefficients one must consider the second order variations in the average energy along the swarm. This is beyond the scope of this work and we defer this to a future study.

Figure 1b clearly shows the anisotropy of the diffusion tensor, i.e. $D_L \neq D_T$; for the rage of $E/n_0$ considered here, the transverse diffusion coefficient is always larger than the longitudinal diffusion coefficient. This is due to the spatial variation of the average energy within the swarm and to the energy dependence of the collision frequency. The theory of anisotropic diffusion in an electric field is now textbook material (see for example \cite{HuxleyC1974}) and rather than a detailed discussion of the origin of this phenomenon, we prefer to highlight the implementation of the diffusion coefficients in fluid models of streamer discharges. For $E/n_0$ below approximately 30 Td, our results for the flux component of the transverse diffusion coefficient and those obtained by the BOLSIG+ code agree very well. For $E/n_0$ above 30 Td, however, BOLSIG+ with its two term approximation is clearly above our multi term solution. Here one should bear in mind that BOLSIG+ treats the diffusion processes under the spatially homogeneous conditions and hence the transverse diffusion coefficient obtained under the spatially-inhomogeneous conditions should be used for comparisons. On the other hand, our one-dimensional fluid model requires the longitudinal diffusion coefficient as an input as we consider the spatial variations of the electron density and average electron energy only along the field direction. Therefore, particular care needs to be taken with implementation of the diffusion coefficients in fluid models of streamer discharges.

Figure 1c shows that the ionization rate differs between the two term and the multi term calculations by up to 30 \%. It is interesting to note that the two term approximation is less accurate in the energy region dominated by the vibrational excitation of N$_2$ and for energies well above the ionization threshold. Surprisingly, for the electric field range between approximately 200 and 600 Td the two term approximation increases in accuracy. In this energy region the cross sections for inelastic processes are much smaller than for elastic collisions. Similar but not identical observations have been made by Phelps and Pitchford \cite{PhelpsP1985}.

\subsection{Input data for the high order model}
\label{sec3d}

\begin{figure} [!htb]
a) \includegraphics[scale=0.3]{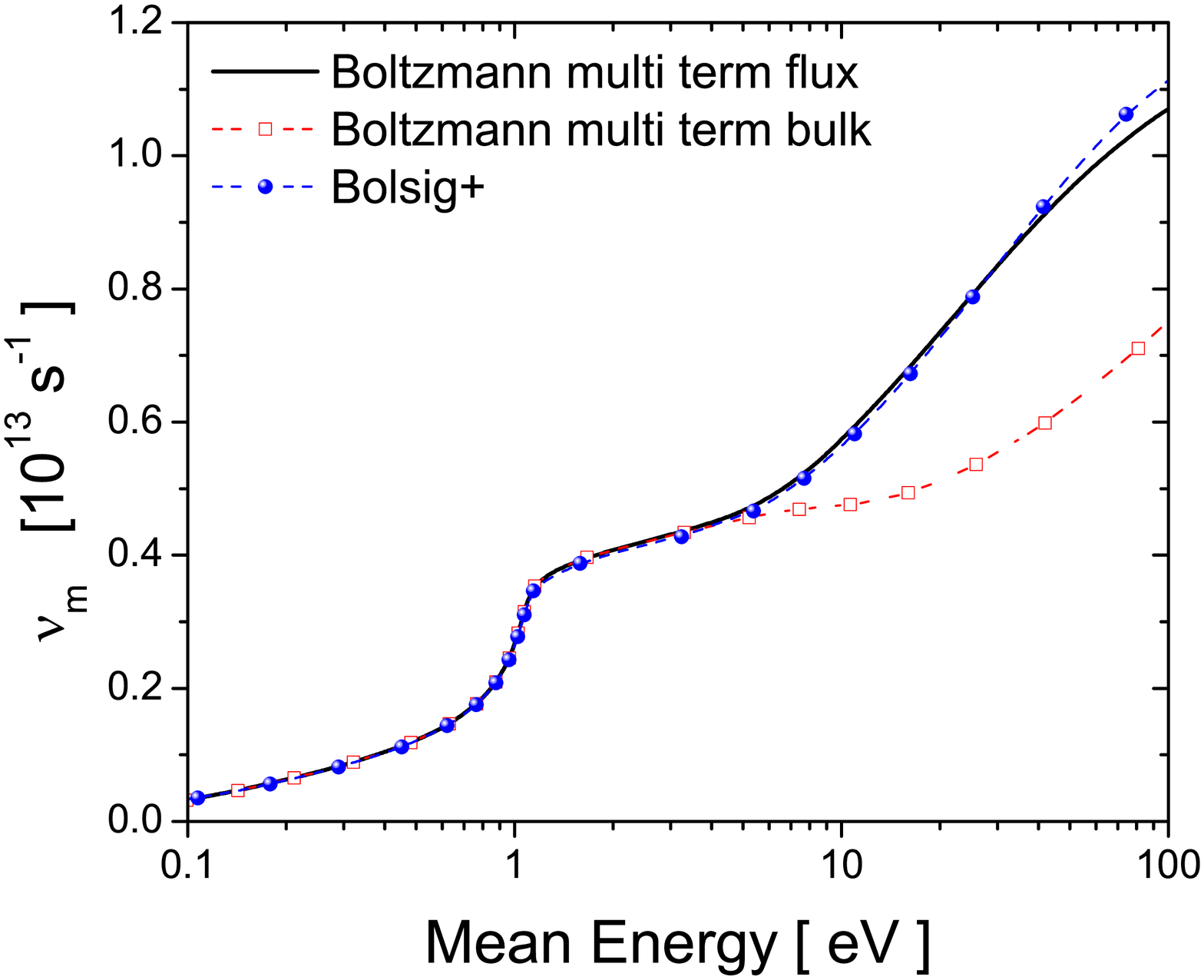}
\label{Fig4}
b) \includegraphics[scale=0.3]{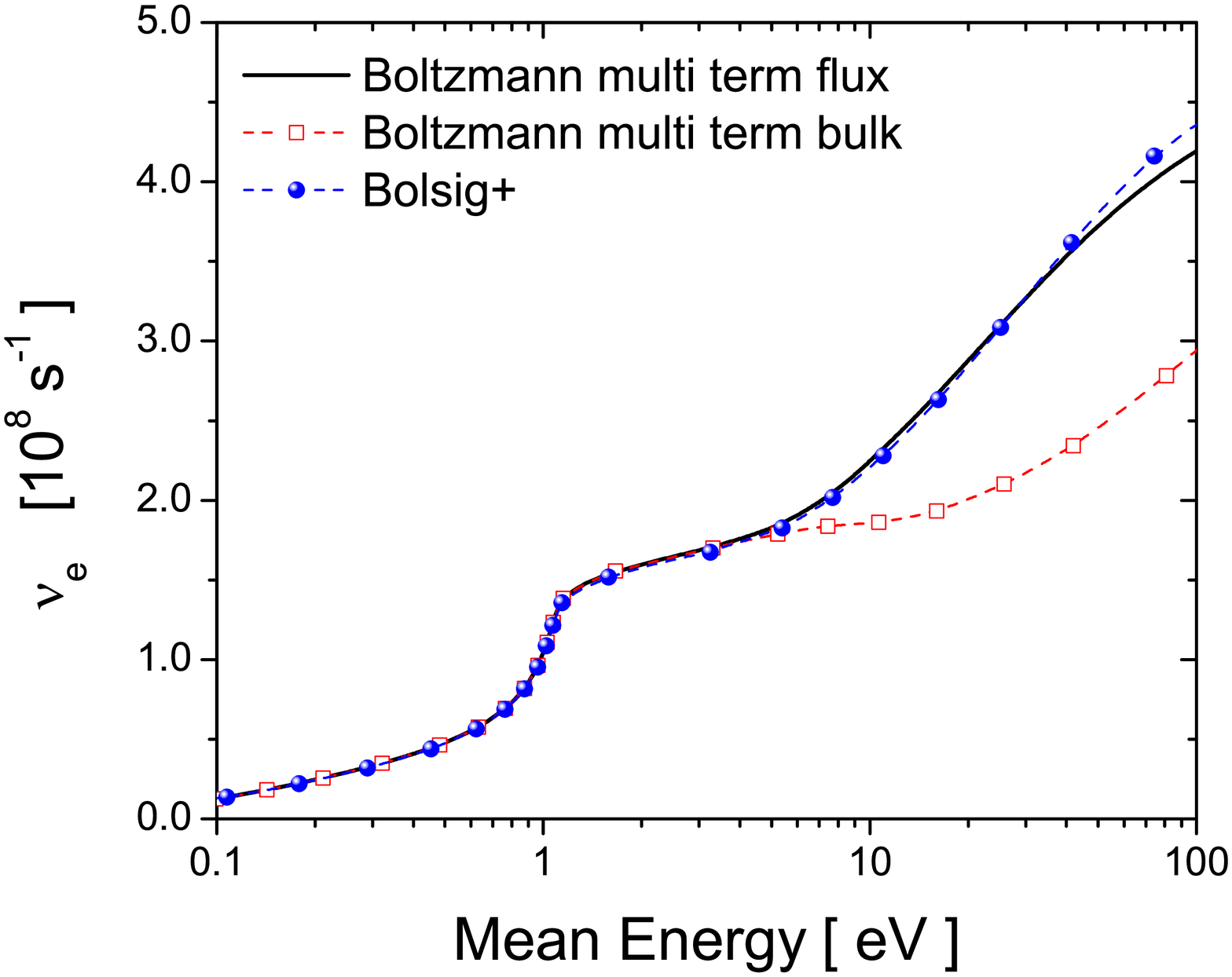}
\label{Fig5}
\caption{Average collision frequency (a) for momentum transfer $\nu_m$, and (b) for energy transfer $\nu_e$ as a function of the mean energy of electrons in N$_2$, as an input for the high order model. The three curves in each panel show the flux and bulk data obtained by our multi term solution for the Boltzmann equation and the flux data obtained by the BOLSIG+ code.}
\end{figure}

\begin{figure} [!htb]
a) \includegraphics[scale=0.3]{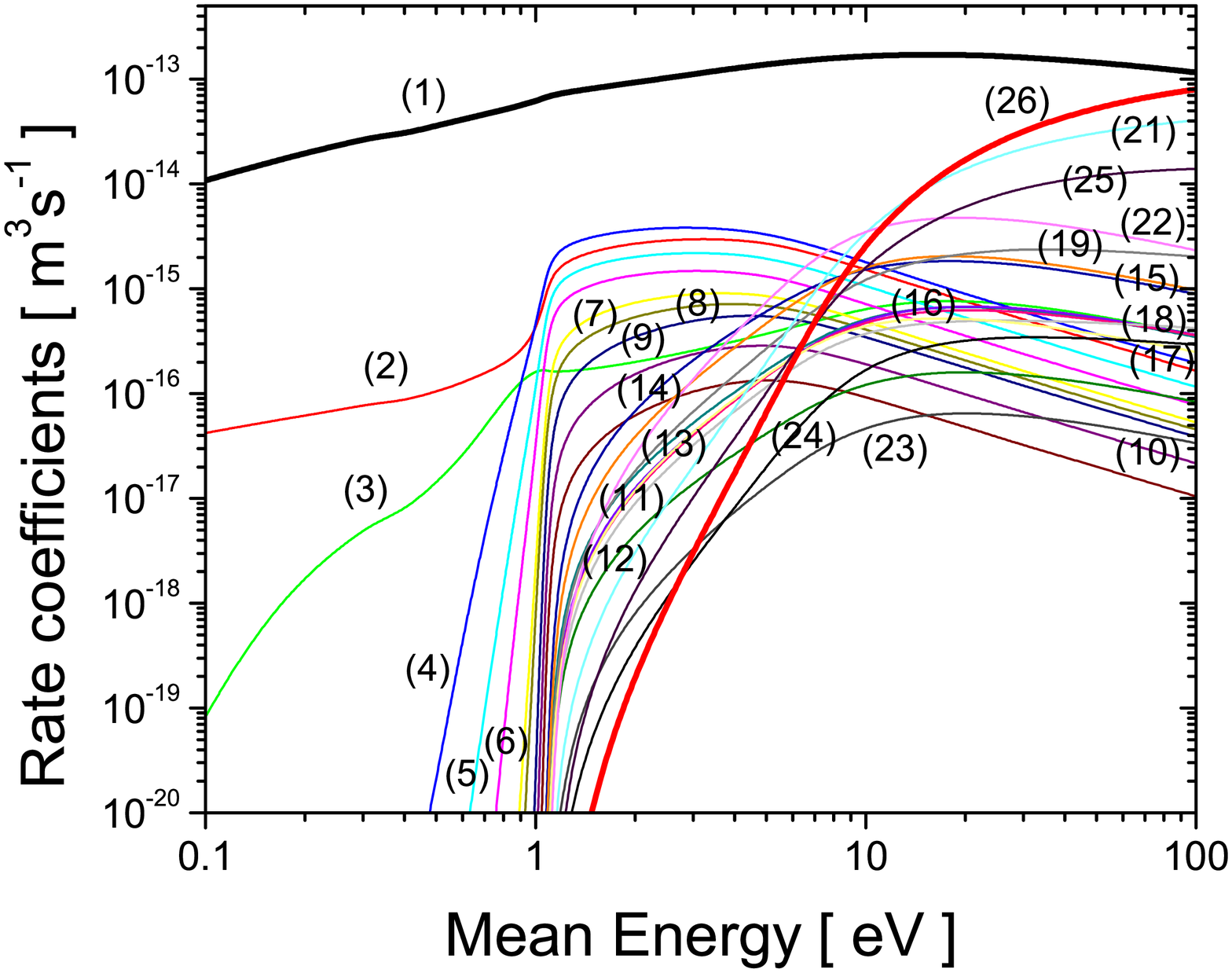}
\label{Fig6}
b) \includegraphics[scale=0.3]{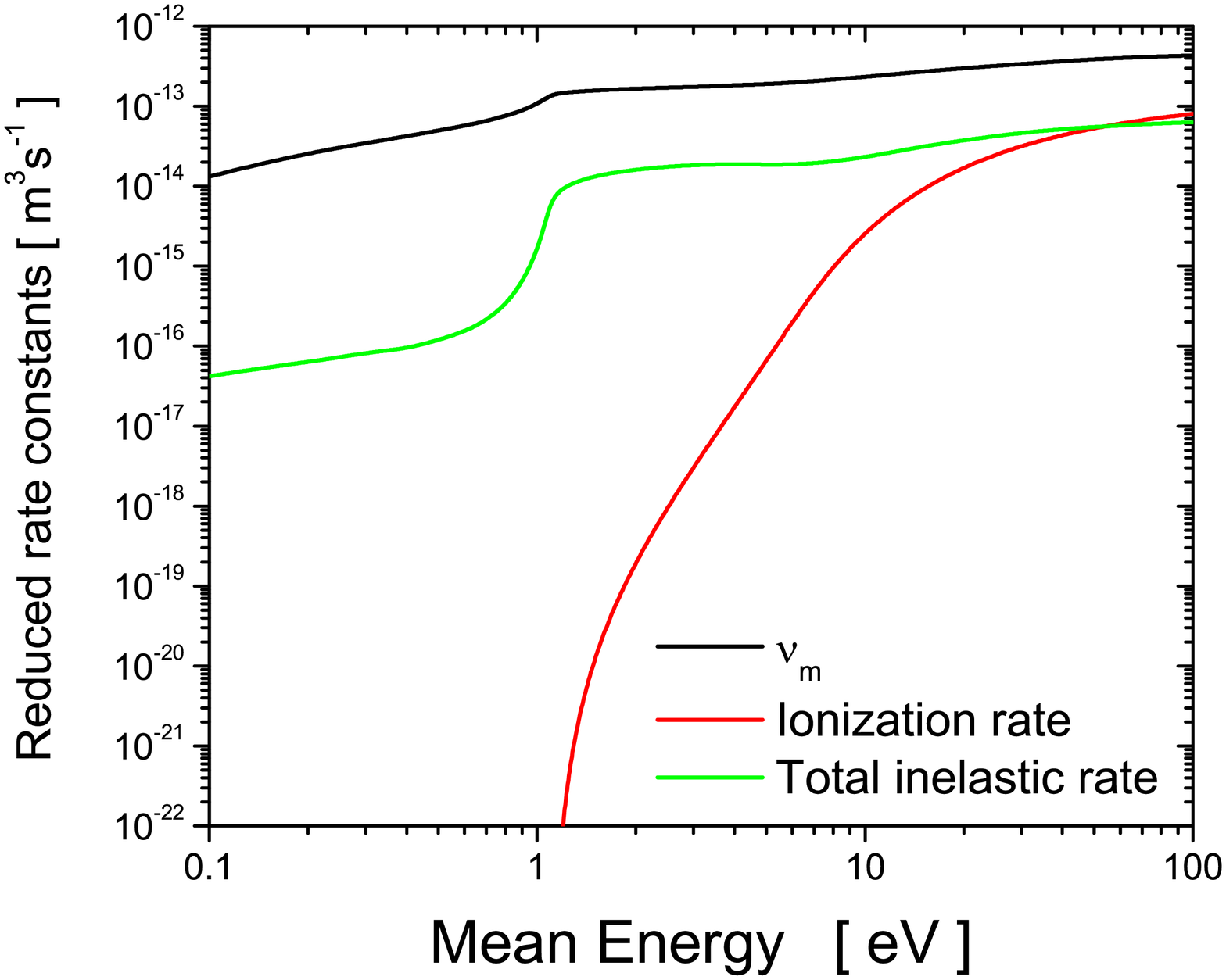}
\caption{Rate coefficients in N$_2$ as a function of the electron energy, calculated with the multi term solution of the Boltzmann equations. (a) Rates for the cross sections listed in~\cite{StojanovicP1998, PhelpsP1985} and (b) momentum transfer rate, total inelastic rate and total ionization rate.}
\label{Fig7}
\end{figure}

The average collision frequencies for momentum and energy transfer as a function of mean electron energy is shown in figure 2. Bulk and flux frequencies obtained with the multi term approach are compared flux values calculated with BOLSIG+. The flux values of the two approaches agree well. The differences between bulk and flux values of the average collision frequencies are a direct consequence of the differences between bulk and flux mobilities (see figure 1a). We see that the average collision frequency for momentum transfer is larger than for energy transfer. This follows from the fact that the energy loss in an elastic collision is proportional to the mass ratio $m/m_0 \ll 1$ (see equation (\ref{2.4.2.10})) and thus much smaller than in an inelastic collision.

Figure 3a shows the rate coefficients as a function of the mean electron energy for all collision processes. Figure 3b compares the momentum transfer rate, the total inelastic rate and the ionization rate. All results in figure 3 are derived with the multi term solution of Boltzmann's equation.
\section{First order streamer model with different transport data}
\label{sec3e}
Here we present simulations of planar negative ionization fronts in N$_2$ using the first order fluid model, while simulation results with the high order fluid model are presented in our second paper~\cite{PaperII}.
\subsection{Numerical methods, initial and boundary conditions.}
\label{sec3f}
In this subsection we briefly describe the numerical method, and the initial and boundary conditions used to solve equations (\ref{2.2.20})-(\ref{2.2.23}) in one spatial dimension. The calculations are carried out in N$_2$ at atmospheric pressure and at the ambient temperature of 298 K. The 1D simulations are started with the same initial Gaussian type distribution for electrons and ions
\begin{equation}
n(x)|_{t=0}=n_i\exp{\Bigg[-\frac{(x-x_0)^2}{\sigma^2}\Bigg]}\,,
\label{3.1}
\end{equation}
in a gap parameterized with the coordinate $x\in[0,L]$, with $L = 1.2$ mm. We have chosen $n_i=2\times10^{18}$ m$^{-3}$, $x_0=8\times10^{-4}$ m and $\sigma=2.9\times10^{-5}$ m. The externally applied electric field is positive in the $x$ direction, therefore electrons drift to the left. The field is fixed in the non-ionized region at the left boundary $x = 0$, providing a fixed electric field for the negative streamer ionization front to penetrate. In this work we consider reduced electric fields of 350 Td, 460 Td, 590 Td, 770 Td and 1000 Td.

To investigate the sensitivity of streamer properties to the definition and accuracy of the transport data, we employ three different sets of data for $\mu$, $D$ and $\nu_I$: bulk and flux data obtained by our Boltzmann equation analysis and the flux data obtained by the BOLSIG+ code.

The finite volume method is used to spatially discretize the system (\ref{2.2.20})-(\ref{2.2.23}) on a uniform grid with 1000 points. More details will be outlined in our second paper~\cite{PaperII} where numerical methods for solving the high order fluid equations are presented in a comprehensive way. To approximate the spatial derivatives in (\ref{2.2.20}) we use the second-order central difference discretization while the time derivatives are approximated with the Runge-Kutta 4 method. The continuity equation for the electron density has a second order spatial derivative, and therefore it requires two boundary conditions. For $x=0$ we use a homogeneous Neumann boundary condition ($\partial_x n =0$), so that electrons that arrive at this boundary may flow out of the system. For $x=L$ we employ a homogeneous Dirichlet boundary condition ($n=0$) to ensure that there is no outflow of electrons from the system. In any case, it should be noted that the electrons are well separated from the boundaries, therefore the actual boundary conditions do not matter. In equation (\ref{2.2.21}) the time derivative is approximated with the Runge-Kutta 4 method using the same time step as in equation (\ref{2.2.20}). In the 1D case, equation (\ref{2.2.23}) has the form
\begin{equation}
\partial_x E = \frac{e}{\epsilon_0} (n-n_{ion})\,,
\label{3.2}
\end{equation}
from which we can determine the electric field $E$ by integrating over $x$ and using the fixed value of $E$ at $x=0$.

\subsection{Overview of simulation results with different transport data}
\label{sec3g}
\begin{figure} [!htb]
a) \includegraphics[scale=0.3]{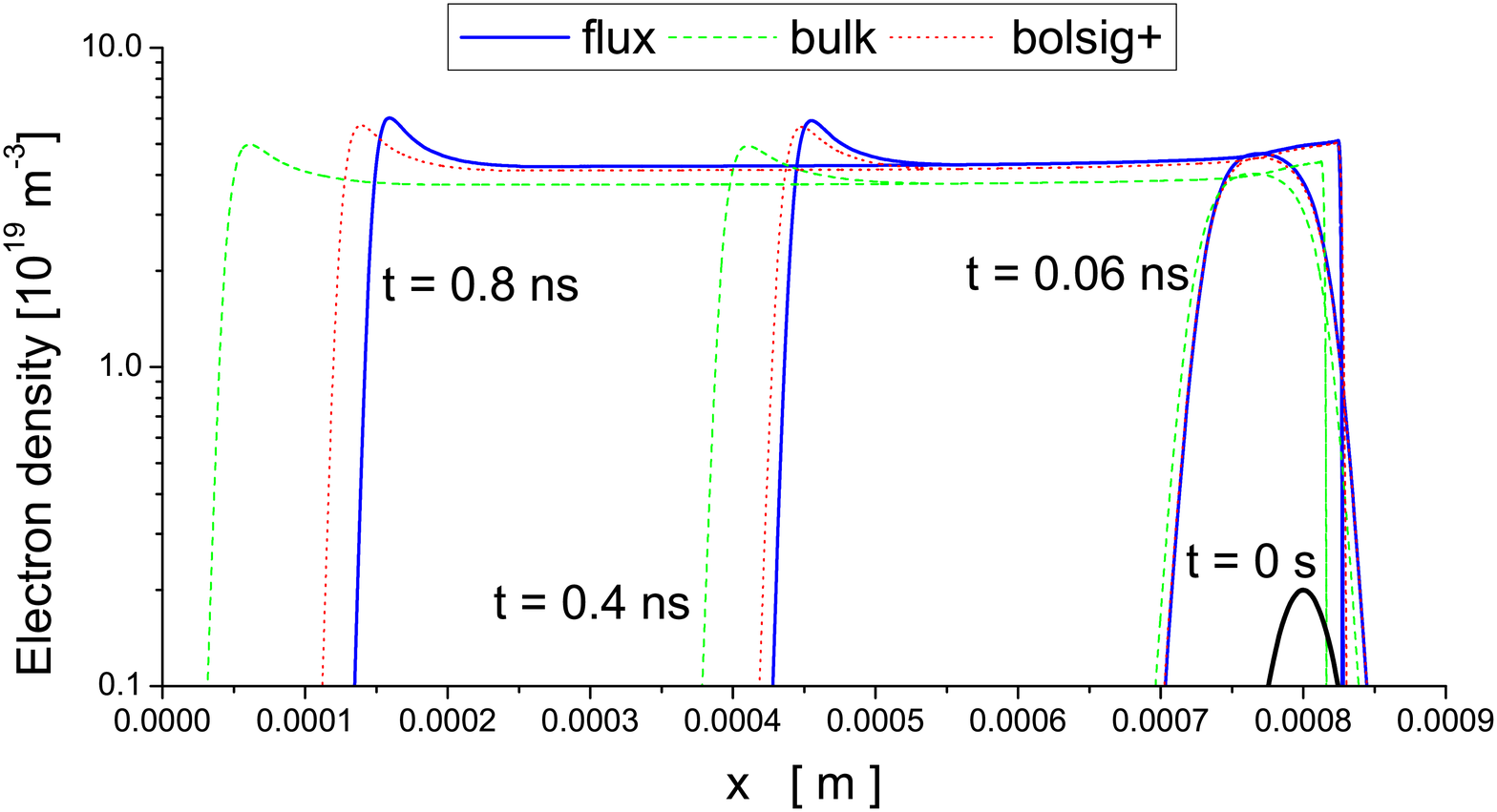}\\
b) \includegraphics[scale=0.3]{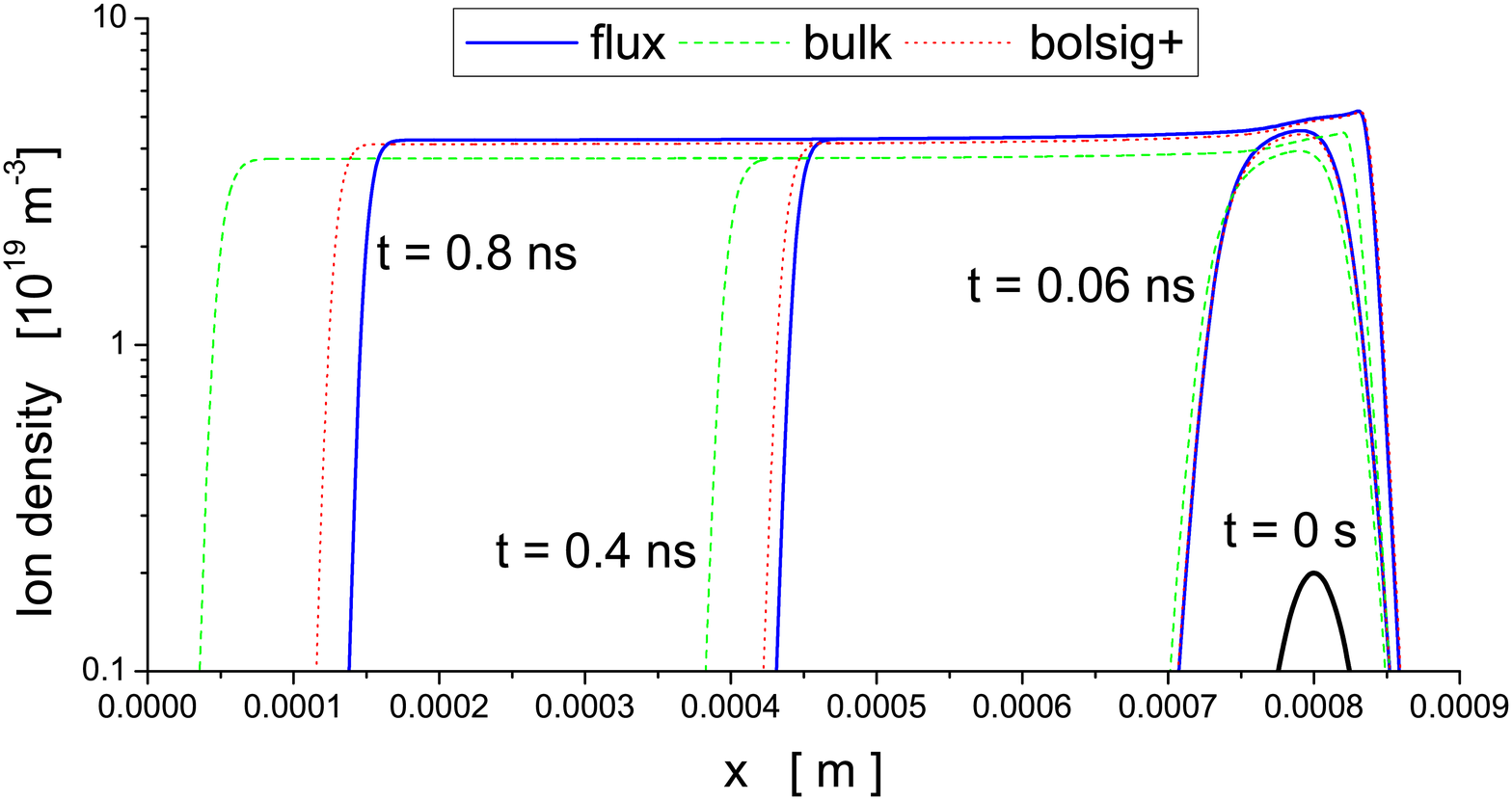}\\
c) \includegraphics[scale=0.3]{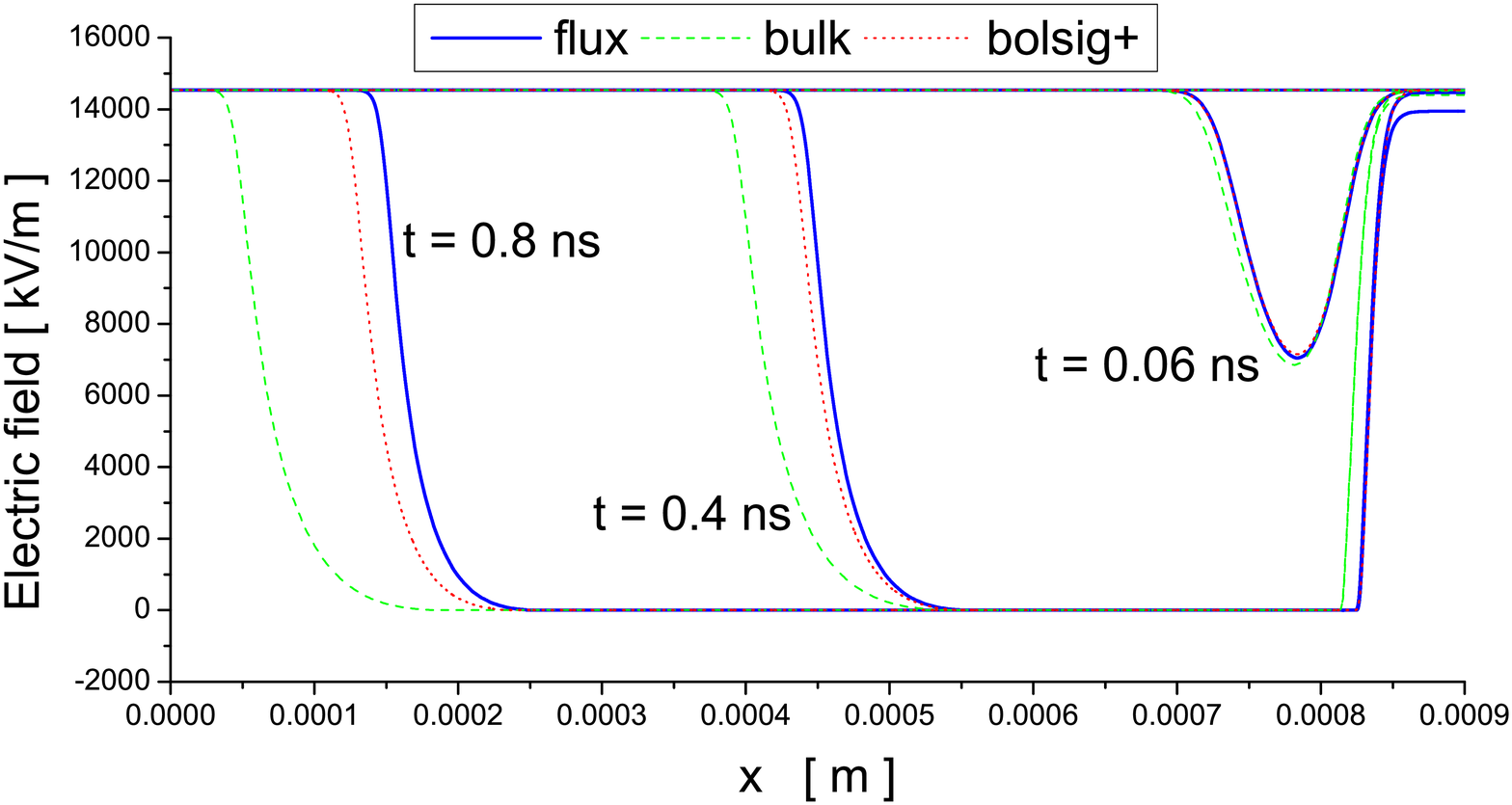}
\caption{Evolution of (a) the electron density, (b) the ion density and (c) the electric field in a negative planar ionization front in N$_2$ with a reduced electric field of 590~Td ahead of the front. Shown are the spatial profiles obtained with three different sets of input data as indicated in the graph.}
\label{Fig8}
\end{figure}

Figure~\ref{Fig8} displays the spatio-temporal evolution of electron and ion densities and of the electric field when the reduced electric field ahead of the front is fixed to 590 Td (or equivalently to 145 kV/cm in N$_2$ at atmospheric pressure and temperature of 298 K). Calculations are performed for three different sets of transport data as indicated in the figures. We start with a Gaussian density distribution as described above. Though the transition from avalanche to streamer has been discussed many times within the past 80 years \cite{LiBEM2007, Raether1964, MontijnE2006}, the characteristics and main physical processes are discussed here to investigate the sensitivity to different transport data.

In the early stage of evolution, we see that both the electron and the ion densities grow due to electron impact ionization. If this was the only mechanism, the electric field would remain unchanged and the ionization would continue indefinitely. However, the electrons drift in the direction opposite to the electric field while the positive ions would slowly drift in the opposite direction; as their mobility is so much smaller, this motion is actually neglected here and in most other streamer studies. As a consequence, the charge separation starts to distort the initially homogeneous electric field. Now figure \ref{Fig8} shows that the ionization profiles at time 0.06 ns obtained with the bulk transport data are somewhat wider while their height is less than with our flux and BOLSIG+ data. This follows from the fact that the bulk mobility and diffusion constant are higher than the corresponding flux data and hence the centre of mass moves faster and the electron package spreads faster. As the ionization rate is the same in both cases, the height of the profiles obtained with the bulk data must be less than with the flux and BOLSIG+ data. As the evolution continues, the electric field in the ionised region gets completely screened, and further ionization processes cannot occur in this region anymore. The transition from avalanche to streamer is then completed. We mention in passing that the complete screening of the interior field is due to the 1D set-up and to the fact, that the field ahead of the front does not change in time.

\subsection{Front velocities}

When the streamers have approached an approximately uniformly translating state, we see that the streamers obtained with bulk data propagate faster than those obtained with flux and BOLSIG+ data, in agreement with earlier studies~\cite{LiTNHE2012, LiBEM2007, LiEH2010}. In section \ref{sec2h} it was already discussed, that in general flux transport data should be used in fluid equations derived systematically from the Boltzmann equation; however, in the particular case of streamer ionization fronts with their pulled dynamics simulated with the half-phenomenological classical streamer model, the bulk coefficients approximate the front velocity better, with the drawback that the ionization level behind the front is too small~\cite{LiTNHE2012, LiBEM2007, LiEH2010}.

\begin{figure} [!htb]
\centering
\includegraphics[scale=0.4]{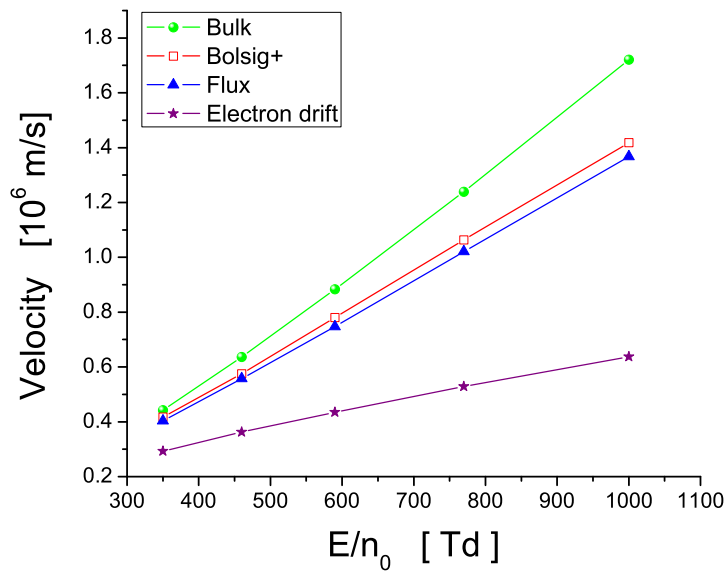}
\caption{Velocities of planar fronts obtained by three different sets of input data as a function of the reduced electric field. The flux drift velocity of electrons is also included.}
\label{Fig11}
\end{figure}

In figure \ref{Fig11} we display velocities of planar fronts as a function of $E/n_0$. We compare the velocities obtained with different input data. The flux drift velocity as a function of $E/n_0$ is also shown. First, we see that the planar fronts move much faster than the electrons. This follows from the fact that the velocity of a planar front is the sum of the drift in the electric field at the front edge plus a term accounting for diffusion, for creation of additional electrons due to impact ionization and for the electron density profile~\cite{EbertSC1997}. The difference between front velocity and electron drift velocity increases with $E/n_0$, up to more than a factor of 2 for the highest field displayed here.

We remark that according to analytical theory~\cite{EbertS2000, EbertSC1997}, the front velocities in planar configurations (where the field does not decay ahead of the front) depend for a long time on initial conditions, and if the initial condition decays less than $e^{-\Lambda^*|x|}$, $\Lambda^*=\sqrt{\nu_I/D}$, it will determine the front velocity for arbitrarily long times. For this reason, we here do not compare numerical with analytical results~\cite{EbertSC1997}.

We finally note that the front velocities calculated with bulk transport data are up to 30 \% higher than with flux data. This illustrates the sensitivity of the model to the input data. On the other hand, the velocities differ much less between our flux data and those obtained with BOLSIG+. To explore this issue in more detail, additional tests are required, particularly for atomic and molecular systems with large anisotropy of the velocity distribution function in velocity space.

\subsection{Ionization levels behind the front}

We now explore how the ionization degree behind the front depends on the transport data, and we compare with the analytical approximation
\begin{equation}
n_{e, \rm back}\leq\frac{\epsilon_0}{e}\int_{0}^{E_{\rm max}} \frac{\nu_I(E)}{E\mu(E)}dE\,.
\label{3.3}
\end{equation}
This approximation becomes an identity, if diffusion can be neglected~\cite{EbertSC1997}; and it is an upper bound, if diffusion is taken into account~\cite{LiBEM2007}. This analytical result has been derived for planar fronts, and it is independent of the front velocity.

\begin{figure} [!htb]
\centering
\includegraphics[scale=0.4]{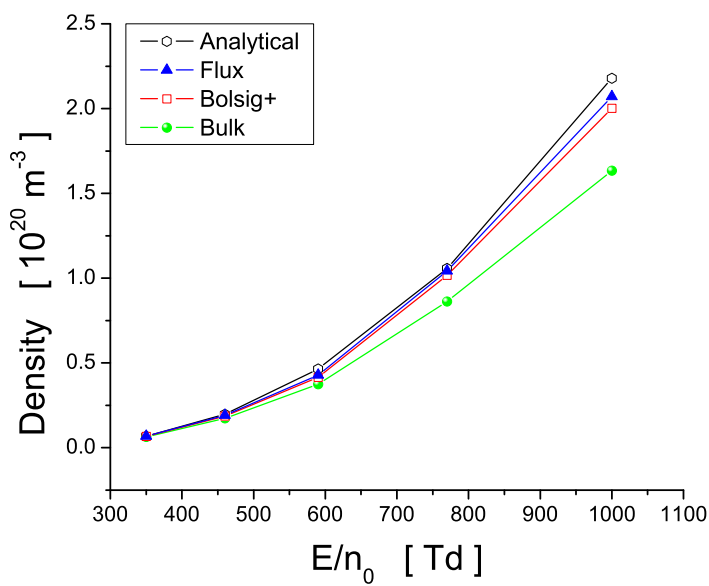}
\caption{Ionization levels behind planar fronts as a function of the reduced electric field. results are shown with three different sets of input data and with the analytical approximation (\ref{3.3}) calculated with flux data.}
\label{Fig12}
\end{figure}

In figure \ref{Fig12} we compare the ionization levels behind the fronts calculated with the three different sets of transport data and with the analytical upper bound, using our flux data. We see that the approximation (\ref{3.3}) indeed serves as such a bound, but is furthermore also a very good approximation of the numerical results, when using the same transport data. The errors of the two term approximation are negligible indicating a weak sensitivity of the ionization level to the isotropy of the distribution function in velocity space that is assumed in BOLSIG+. The ionization level with bulk data is considerably lower. Generally, it is evident that the ionization level is much less sensitive to the type of transport data than the front velocity. Similar observations have been made in~\cite{LiBEM2007}.
\section{Conclusion}
In this paper we have derived a high order fluid model for streamer discharges. Our goal has been to develop a comprehensive theory at a level of sophistication appropriate for the electron energy distributions very far from equilibrium and for the steep electron density gradients that characterize a streamer front. The first steps are reported in the present paper, which deals with theoretical foundations and phenomenology. After examining the application of the Boltzmann equation, we have proceeded to the evaluation of the velocity moments of Boltzmann's equation. To make contact with previous works and to place our theory in a much broader context, we have demonstrated how to derive the classical drift-diffusion approximation often used in the plasma modeling community to model streamer discharges. Then we have introduced our more sophisticated high order fluid approach, proceeding directly from Boltzmann's equation and systematically discussing the critical assumptions required to close the system of equations and to evaluate the collision terms involved. Momentum transfer theory has been used to approximate collision terms in the high order fluid model while high order tensors appearing in the energy flux equation have been specified in terms of previous moments. In contrast to previous works, it has been emphasized that the energy flux equation plays a pivotal role for the correct description of streamer dynamics. The fluid equations obtained as velocity moments of the Boltzmann equation have been closed in the local mean energy approximation and coupled to the Poisson equation to calculate the modification of the electric field by space charges. The numerical solutions of the high order fluid model for planar fronts and their discussion are deferred to the following paper.

The second important aspect of this paper concerns the application of transport data in fluid models of streamer discharges. In order to illustrate this issue, we have used the first order fluid model to investigate the temporal evolution of negative planar fronts in pure nitrogen. We have focused on the way in which the inherent streamer properties such as the velocity of a streamer or the ionization level behind the front are influenced by different transport data employed as an input in fluid equations. Our primary goal was to show which aspects of kinetic theory developed for swarm physics and particularly which segments of data would be important for further improvement of streamer models. It was shown and illustrated that the direct application of transport data from the literature without knowledge of origin and nature of the data is problematic and can often lead to significant errors in the profiles of various streamer properties. In this respect, the origin and nature of transport data must be known and, if appropriate, suitably modified before implementation in the fluid models. This is particularly important for collisional transfer rates required as input in high order fluid model. We have also discussed the validity of transport data obtained by a two term theory for solving the Boltzmann equation. Our general sentiment was that two term data may be used for fluid modeling of streamers only if demands for high accuracy are relaxed.

\ack{It is a pleasure to acknowledge very helpful discussions with Robert Robson, Willem Hundsdorfer and Chao Li. Furthermore, SD and AM acknowledge support from STW-projects 10118 and 10751, part of the Netherlands's Organization for Scientific Research (NWO). SD has also been supported by the MNTR, Serbia, under the contract number ON171037. RDW is supported by the Australian Research Council and the Centre for Antimatter-Matter Studies, Australia.}
\begin{appendix}
\section{Alternative forms of the high order model} \label{app}

To make contact with previous work \cite{KanzariYH1998, AbbasB1981, BayleC1985, EichwaldDMYD2006}, we here present different forms of equations (\ref{2.3.1})-(\ref{2.3.4}). When we evaluate the averages over the velocities $\vc c$ in these equations and perform a considerable amount of algebra, we find
\begin{equation}
\fl
\frac{\partial n}{\partial t} + \nabla\cdot n\vc{v} = C_1,
\label{2.3.11}
\end{equation}
\begin{equation}
\fl
\frac{\partial}{\partial t}\left( nm\vc{v}\right) + \nabla\cdot\left( nm\vc{v}\vc{v}\right)+ \nabla\cdot\vc{P} - ne\vc{E} = C_{m\bf{c}},
\label{2.3.12}
\end{equation}
\begin{equation}
\fl
\frac{\partial }{\partial t}\left[n\left( \frac{1}{2}mv^2+\frac{3}{2}kT \right)\right] +\nabla\cdot\left[\left(\frac{1}{2}mv^2+\frac{3}{2}kT\right)n\vc{v} + \vc{P}\cdot{\vc{v}}+\vc{q}\right]-ne\vc{E}\cdot\vc{v} = C_{\frac{1}{2}mc^2},
\label{2.3.13}
\end{equation}
\begin{eqnarray}
\fl
\frac{\partial }{\partial t}\left[ \left(\frac{1}{2}mv^2+\frac{3}{2}kT\right)n\vc{v} + \vc{P}\cdot{\vc{v}} + \vc{q} \right] +\nabla\cdot\Bigg[ 2\vc{v}\left(\vc{P}\cdot\vc{v}\right) + 2\vc{v}\vc{q} +\frac{1}{2}v^2\vc{P}+\vc{Q}:\vc{v} + \vc{S}\nonumber \\
\fl
+\left(\frac{1}{2}mv^2 + \frac{3}{2}kT\right)n\vc{v}\vc{v}\Bigg]-\frac{e}{m}\left( \frac{1}{2}mv^2+\frac{3}{2}kT \right)n\vc{E}- en(\vc{v}\vc{v})\cdot\vc{E} \nonumber \\
\fl
-\frac{en}{m}\vc{P}\cdot\vc{E} = C_{\frac{1}{2}mc^2\bf{c}},
\label{2.3.14}
\end{eqnarray}
where $\vc{P}$ is the pressure tensor, $\vc{q}=\frac{1}{2}nm\big\langle\left(\vc{c}-\vc{v}\right)^2(\vc{c}-\vc{v})\big\rangle$ is the heat flux vector, $\vc{S}=\frac{1}{2}nm \big\langle\left(\vc{c}-\vc{v}\right)^2(\vc{c}-\vc{v})(\vc{c}-\vc{v})\big\rangle$ is the high order pressure tensor and $\vc{Q}=nm\big\langle (\vc{c}-\vc{v})(\vc{c}-\vc{v})(\vc{c}-\vc{v})\big\rangle$ is the high order heat flux tensor. The pressure tensors $\vc{P}$ and $\vc{S}$ are the second order tensors while the high order energy flux tensor $\vc{Q}$ is a third order tensor.

Using the continuity equation (\ref{2.3.11}), the momentum balance equation (\ref{2.3.12}) can be written as
\begin{equation}
nm\left[\frac{\partial \vc{v}}{\partial t} + \left(\vc{v}\cdot\nabla\right)\vc{v}\right] - \nabla\cdot\vc{P}-en\vc{E} = C_{m\bf{c}} - m\vc{v}C_1.
\label{2.3.15}
\end{equation}
This equation is equivalent to the equation (\ref{2.2.9}) and can be used to exclude the energy of macroscopic motion from the energy (\ref{2.3.13}) and energy flux (\ref{2.3.14}) balance equations, respectively. For this purpose, we multiply momentum balance equations (\ref{2.3.12}) and (\ref{2.3.15}) by $\vc{v}$ and after addition of one of the resulting equations to another one, we obtain
\begin{equation}
\frac{\partial}{\partial t}\left( \frac{1}{2}nmv^2\right)+\nabla\cdot\left( \frac{1}{2}nmv^2\vc{v}\right) + (\nabla\cdot\vc{P})\cdot\vc{v} = \vc{v}C_{m\bf{c}} - \frac{1}{2}v^2C_1.
\label{2.3.16}
\end{equation}
This is the balance equation for the energy of macroscopic motion. Taking this equation back into the energy and energy flux balance equations, we obtain an alternative and more compact form of the fluid equations
\begin{equation}
\fl
\frac{\partial n}{\partial t} + \nabla\cdot n\vc{v} = C_1,
\label{2.3.5}
\end{equation}
\begin{equation}
\fl
nm\frac{d\vc{v}}{dt} +\nabla\cdot\vc{P} -en\vc{E} = C_{m\bf{v}_r},
\label{2.3.6}
\end{equation}
\begin{equation}
\fl
\frac{d}{dt}\Big(\frac{3}{2}p\Big) +\frac{3}{2}p(\nabla\cdot\vc{v})+\nabla\cdot\vc{q}+\vc{P}:\nabla\vc{v} = C_{\frac{1}{2}mv_r^2},
\label{2.3.7}
\end{equation}
\begin{equation}
\fl
\frac{d\vc{q}}{dt}+\vc{q}\cdot\nabla\vc{v}+\vc{q}(\nabla\cdot\vc{v})+\vc{Q}:\nabla\vc{v}+\nabla\cdot\vc{S}
+\Big(\frac{d\vc{v}}{dt}-\frac{e}{m}\vc{E}\Big)\Big(\vc{\tau}+\frac{5}{2}p\vc{I}\Big)= C_{\frac{1}{2}mv_r^2\bf{v}_r}.
\label{2.3.8}
\end{equation}
where $\vc{v}_r=\vc{c}-\vc{v}$ is the random velocity. The pressure $p$ is defined as one-third the trace of the pressure tensor
\begin{equation}
p 
=\frac{1}{3}\sum_{i}P_{ii} = \frac{m}{3}\int \left(\vc{c}-\vc{v}\right)^2fd\vc{c}\,,
\label{2.3.9}
\end{equation}
while $\vc{\tau}$ is the stress tensor
\begin{equation}
\vc{\tau}=\vc{P}-p\vc{I}\,,
\label{2.3.10}
\end{equation}
where $\vc{I}$ is the unity tensor (diagonal elements equal to unity). In the equations (\ref{2.3.6})-(\ref{2.3.8}), the convective time derivative (\ref{2.2.8}) has been utilized. The collisional terms on the right-hand-side of the balance equations (\ref{2.3.6})-(\ref{2.3.8}) are given by
\begin{eqnarray}
C_{m\bf{v}_r} = -\int m\vc{v}_rJ(f)d\vc{v}_r, \\
C_{\frac{1}{2}mv_r^2} = - \int \frac{1}{2}mv_r^2J(f)d\vc{v}_r, \\
C_{\frac{1}{2}mv_r^2\bf{v}_r} = - \int \frac{1}{2}mv_r^2\vc{v}_rJ(f)d\vc{v}_r,
\end{eqnarray}
where $J(f)$ is the collision operator and $f$ is the distribution function of charged particles. It should be noted that balance equations (\ref{2.3.5})-(\ref{2.3.8}) can be derived directly from the Boltzmann equation. In such case, however, the Boltzmann equation must be transformed into an equation for $\vc{v}_r$ before the velocity moments are taken.

\end{appendix}

\end{document}